\documentclass[pdflatex,sn-mathphys-num]{sn-jnl}

\usepackage{graphicx}%
\usepackage{multirow}%
\usepackage{amsmath,amssymb,amsfonts}%
\usepackage{amsthm}%
\usepackage{mathrsfs}%
\usepackage[title]{appendix}%
\usepackage{xcolor}%
\usepackage{textcomp}%
\usepackage{manyfoot}%
\usepackage{booktabs}%
\usepackage{algorithm}%
\usepackage{algorithmicx}%
\usepackage{algpseudocode}%
\usepackage{listings}%
\usepackage{comment}
\usepackage{tabularx}
\usepackage{array}
\usepackage{booktabs}

\theoremstyle{thmstyleone}%
\theoremstyle{thmstyletwo}%

\theoremstyle{thmstylethree}%

\begin{document}

\title[Article Title]{AoNT Trap: Borromean-Entangled Mutable
Chameleon Trapdoor Hash
All-or-Nothing Stream Cipher}


\author[]{\fnm{Victor} \sur{Kebande}}\email{victor.kebande@ucdenver.edu}

\affil{
\orgdiv{Department of Computer Science}

\orgname{University of Colorado Denver}

\orgaddress{
\street{1380 Lawrence Street},
\city{Denver},
\state{Colorado},
\postcode{80204},
\country{USA}}}


\abstract{This work introduces the Borromean-Entangled Chameleon Trapdoor Hash   All-or-Nothing (\texttt{AoNT}) Stream Cipher (\texttt{BEC-Trap}), a novel construction that merges Borromean interdependence, trapdoor-enabled mutability, and streaming encryption into a unified framework. The (\texttt{BEC-Trap}) cipher links key (\(K\)), initialization vector (\(V\)), and internal state (\(S_t\)) in a Borromean structure, ensuring that breaking, guessing, or removing any one component collapses the entire keystream, providing a computational (\texttt{AoNT}) interdependence under standard cryptographic assumptions. A chameleon trapdoor hash is integrated to permit controlled collisions, enabling seamless rekeying, (\(V\)) refresh, and state rotation without resynchronizing endpoints. This design provides confidentiality, forward secrecy, and adaptive key management with low computational overhead, making it suitable for high-throughput secure messaging, IoT communications, and privacy-preserving blockchain channels. Security analysis of the (\texttt{BEC-Trap}) shows that the construction is resistant to key-recovery attacks, state compromise, and desynchronization attempts, delivering a robust cryptographic primitive for next-generation secure communications.}




\maketitle

 \section{Introduction}
Continued evolution of the stream ciphers as fundamental primitives in modern symmetric cryptography has seen them prized for their high throughput, low latency, and efficient implementation in hardware and software environments\cite{jiao2020stream,manifavas2016survey}. In contrast to the block ciphers that operate on fixed-size data units, the stream ciphers are able to produce a continuous pseudo-random keystream that is combined with plaintext using simple operations such as XOR \cite{klapper2010pseudorandom}. This continuous pseudo-random keystream generation property enables the stream ciphers to encrypt data on-the-fly \cite{kuznetsov20249}. As a result, this   makes them ideal for environments that need real-time responsiveness like wireless communications \cite{kuznetsov2024high}, embedded systems, and Internet-of-Things (IoT) devices \cite{manifavas2016survey}.  

However, despite the stream cipher's efficiency, traditional stream ciphers suffer from critical limitations concerning key management, synchronization, and resilience to partial compromise. It has been observed that once the stream ciphers are initialized, their internal states are able to evolve deterministically, and as such, a leakage or corruption of the key \texttt{ (\(K\))}, Initialization Vector\texttt (\(V\)), or Internal State \texttt (\(S_t\)) often results in catastrophic failure or complete exposure of the keystream.

There exist several well-known stream cipher families, like Chacha \cite{bernstein2008chacha}, ChaCha20 \cite{de2017chacha20}, \texttt{EChaCha20} \cite{kebande2023extended} Salsa20 \cite{bernstein2008salsa20}, Grain\cite{hell2007grain}, and Trivium \cite{de2006trivium}, that have achieved significant success in speed and diffusion strength. However, their architectures typically treat the \texttt{\(K\)}, \texttt{\(V\)}, and \texttt{\(S_t\)} as distinct entities with limited interdependence. This structural separation introduces potential security risks. For example,  an adversary that is capable of isolating or predicting one of the components (\texttt{key}, \texttt{IV} or the \texttt{internal state}) component may exploit it to infer partial keystreams, mount state-recovery attacks, or cause desynchronization between communicating endpoints. In addition to this, conventional rekeying mechanisms rely on session restarts or state regeneration. This in the long run  incurs synchronization costs and increase s the vulnerability window in dynamic communication systems.

Given the  evolving cyber threat  landscape particularly in secure communication, decentralized and adaptive ecosystems, there is a need for  a more cohesive secure-centric  paradigm that unifies interdependence, adaptability, and cryptographic accountability. To this end, we introduce a fundamentally new stream cipher architecture that integrates these properties into a unified framework herein referred to as  Borromean-Entangled Chameleon Trapdoor Hash All-or-Nothing Stream Cipher (\texttt{BEC-Trap}).

The design of \texttt{BEC-Trap} draws its inspiration from three distinct yet complementary cryptographic and mathematical constructs. First, the we leverage the Borromean ring model \cite{cromwell1998borromean}, which  introduces structural interdependence among the (\texttt{key}, \texttt{IV} and the \texttt{cipher state}). Second a chameleon trapdoor hash mechanism provides controlled mutability and authorized collision generation.   Third, an all-or-nothing, (\texttt{AoNT}) security property \cite{rivest1997all},  provides a computational  interdependence under standard cryptographic assumptions. It guarantees that partial exposure of the cipher’s internal components yields no meaningful information about the keystream or plaintext. 

\noindent The contributions of this paper are as follows:
\begin{itemize}
    \item We propose \texttt{BEC-Trap}, a novel stream cipher architecture that unifies Borromean interdependence, trapdoor-based mutability, and all-or-nothing (\texttt{AoNT}) security within a single cryptographic framework.
    \item We introduce a \texttt{Borromean linkage model} a  that mathematically couples the (\texttt{key}, \texttt{IV} and the \texttt{internal state}), ensuring total interdependence such that the compromise of any element disrupts the entire keystream.
    \item We integrate a \texttt{chameleon trapdoor hash layer} that enables controlled collisions for authorized rekeying and state refresh without requiring full session reinitialization or desynchronization.
    \item We conduct a comprehensive security and performance analysis, demonstrating robustness against key-recovery, exposure and state-compromise, and resistance to attacks while maintaining low computational overhead.
\end{itemize}
The remainder of this paper is organized as follows: Section II discuss the Background of \texttt{Borromean Rings}'s and \texttt{Trapdoor Hash }  foundations. This is then followed by   Section III  and IV that discusses the Related Work and Research Gap respectively. After this, the Threat Model and the  Methodology are discussed in Section V and VI respectively. Next, the Experiments are discussed in Section VII followed by the Results in Section VIII. An Security and Perfomance Analysis is given in Section IX and X respectively followed by a Theoretical Validation in Section XI. The paper ends with a Conclusion and a mention of  Future Work in Section XII.

\section{Background}

\subsection{ Borromean Rings}

The Borromean rings are a classical object in topology and knot theory, consisting of three interlocked rings that form a single connected structure, even though no two rings are directly linked to one another \cite{erhardt1997borromean}. If any one ring is removed, the remaining two become completely unlinked. This configuration illustrates a fundamental principle of interdependence and nontrivial linking, serving as a canonical example of a Brunnian link, which is a link that becomes disconnected when any component is removed \cite{penney1969generalized}.

The Borromean rings derive their name from the Italian Borromeo family, whose coat of arms from the Renaissance period prominently featured three interlaced rings as a symbol of strength through unity \cite{jablan1999borromean}. Over time, this configuration has appeared in diverse contexts, from medieval art and religious symbolism to modern physics and mathematics, representing harmony, balance, and mutual dependence.

\begin{figure}[H]
\centering
\includegraphics[width=7cm]{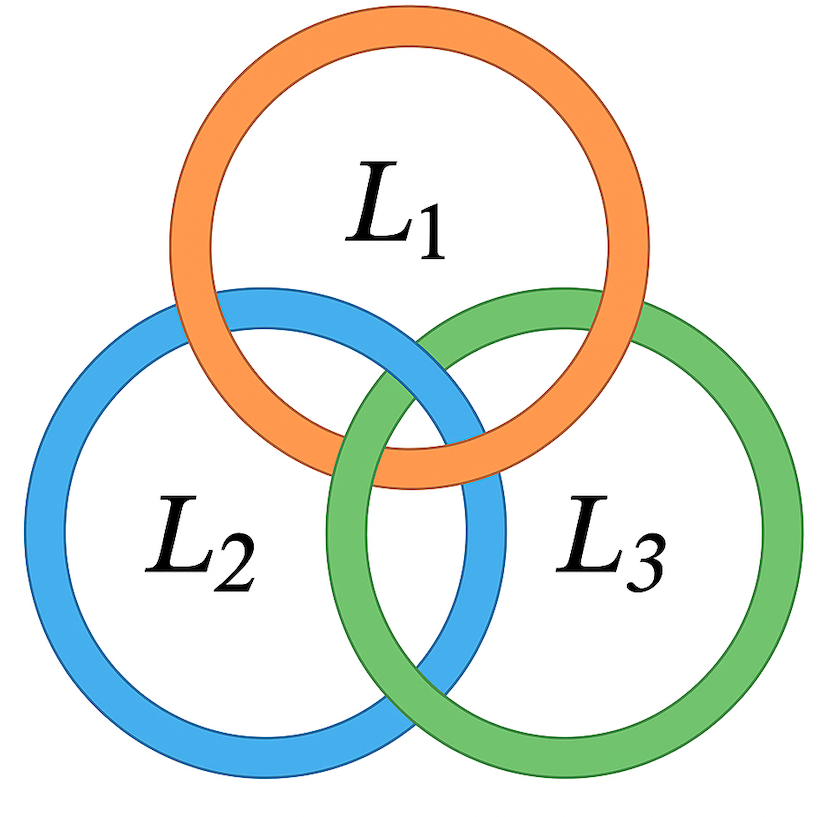}
\caption{Borromean rings represented as three interlocked loops \(L_1\), \(L_2\), and \(L_3\) in \(\mathbb{R}^3\). Each ring is individually unlinked from the others, yet the entire configuration forms a nontrivial link that becomes disconnected upon the removal of any one component.}
\label{KPES}
\end{figure}

From a mathematical standpoint, the Borromean rings can be rigorously described using concepts from algebraic topology \cite{guan2025topological}. From Figure \ref{KPES}, it is evident that each ring  is shown from Figure \ref{KPES}, \(L_i\) can be viewed as an embedding of the circle \(S^1\) into three-dimensional Euclidean space, such that the entire configuration \(L = (L_1, L_2, L_3)\) forms a three-component link in \(\mathbb{R}^3\) as is shown in Equation \ref{eq:linking}. The remarkable feature of this link is that all pairwise linking numbers vanish:
\begin{equation}
\text{Link}(L_i, L_j) = 0, \quad \forall\, i \neq j
\label{eq:linking}
\end{equation}

yet the total link \(L\) is nontrivial. This property implies that no two rings are directly coupled \cite{erhardt1997borromean}, but the overall system cannot be decomposed into disjoint parts without severing at least one component. Important to note is that, the linkage is preserved not through pairwise entanglement but through the collective spatial embedding of all three components.

Consequently, from  a topological perspective, the Borromean rings are often used to illustrate higher-order linking phenomena \cite{guan2025topological}. Their complement in three-dimensional space has a hyperbolic structure \cite{wielenberg1981three} \cite{hilden1992borromean}, and they serve as one of the simplest examples of a nontrivial link with vanishing pairwise linking numbers. The minimal crossing number of the Borromean rings is six, and the configuration has been extensively studied in knot theory as an example of nontrivial linking with trivial pairwise interactions. In addition, the Borromean structure embodies the same fundamental principle: stability and coherence arise not from pairwise bonds but from the interdependence of all components in the system.

Furthermore, each component \(L_i\) that is shown in Figure \ref{KPES} of the Borromean rings can be parametrically represented using trigonometric embeddings in toroidal coordinates, ensuring rotational symmetry and equal geometric constraints. These parameterizations reveal that the rings occupy distinct, yet interdependent, planes in three-dimensional space, maintaining perfect symmetry about the origin. Removing any one component causes the system’s embedding to collapse into three disjoint circles, thereby confirming the topological rule that the integrity of the Borromean structure depends on the simultaneous existence of \(L_1\), \(L_2\), and \(L_3\).

\subsection{Trapdoor Hash Foundations}

Trapdoor hash functions \texttt{THF} are a class of cryptographic primitives that are able to combine the one-wayness of conventional hash functions, with the selective reversibility enabled by a hidden secret \cite{catalano2015algebraic}, referred to as the \textit{trapdoor}. The \texttt{THF} are designed as follows: For any given input, the hash output can be efficiently computed by all users, but only the possessor of the trapdoor can generate a controlled collision that is, a distinct input that produces the same hash value \cite{chandrasekhar2012trapdoor}. This capability introduces asymmetry: the function remains collision-resistant for everyone except the authorized entity in possession of the trapdoor.

Formally, we let \(\mathcal{H}\) denote a family of hash functions parameterized by a public key \(pk\) and a secret trapdoor \(t\). The \texttt{THF} system is defined by three algorithms:
\[
(\textsf{Gen}, \textsf{Hash}, \textsf{Collide})
\]
where:
\begin{itemize}
    \item \(\textsf{Gen}(1^\lambda)\) is a probabilistic key generation algorithm that, given a security parameter \(\lambda\), outputs a pair \((pk, t)\), where \(pk\) are public parameters and \(t\) is the trapdoor.
    \item \(\textsf{Hash}_{pk}(m, r)\) is a public algorithm that maps a message \(m\) and random nonce \(r\) to a digest \(h\), such that \(h = \textsf{Hash}_{pk}(m, r)\).
    \item \(\textsf{Collide}_t(m, r, m')\) is a deterministic algorithm that, given \(t\), computes a value \(r'\) such that:
    \begin{equation}
\textsf{Hash}_{pk}(m, r) = \textsf{Hash}_{pk}(m', r')
\label{eq:collision}
\end{equation}

\end{itemize}
For any entity lacking the trapdoor \(t\), finding such a pair \((m', r')\) is computationally infeasible as is shown in Equation \ref{eq:collision}, assuming standard hardness assumptions such as the discrete logarithm problem or integer factorization. This construction ensures that while the hash function behaves like a conventional one-way function for the public, it offers controlled mutability for authorized entities.

Trapdoor hashes are built upon \textit{trapdoor one-way functions} \cite{bhowmik2024unorthodox}, which are functions \(f(x)\) that are easy to compute but hard to invert without special knowledge. A simple example is an RSA-based trapdoor hash. We let \(n = pq\) be the product of two large primes, and let \(e\) and \(d\) be public and private exponents satisfying \(ed \equiv 1 \pmod{\phi(n)}\) \cite{catalano2002hardness}. The public hash may be defined as is shown in Equation \ref{eq:trapdoor-hash}:
\begin{equation}
h = (g^{m} \cdot r^{e}) \bmod n
\label{eq:trapdoor-hash}
\end{equation}

where \(g\) and \(r\) are elements of \(\mathbb{Z}_n^*\). Without knowing \(d\), finding another pair \((m', r')\) that yields the same \(h\) requires solving the discrete logarithm modulo \(n\), which is computationally infeasible \cite{bhowmik2024unorthodox,haastad2004security}. However, the trapdoor holder (who knows \(d\)) can compute \(r'\), \cite{chandrasekhar2012trapdoor} (Equation \ref{eq:collision-generation})  such that:
\begin{equation}
r' = r \cdot (g^{m - m'})^{d} \bmod n
\label{eq:collision-generation}
\end{equation}

thereby generating a valid collision in constant time. This mathematical asymmetry underlies the core security property of trapdoor hash functions.

\begin{figure}[H]
\centering
\includegraphics[width=8cm]{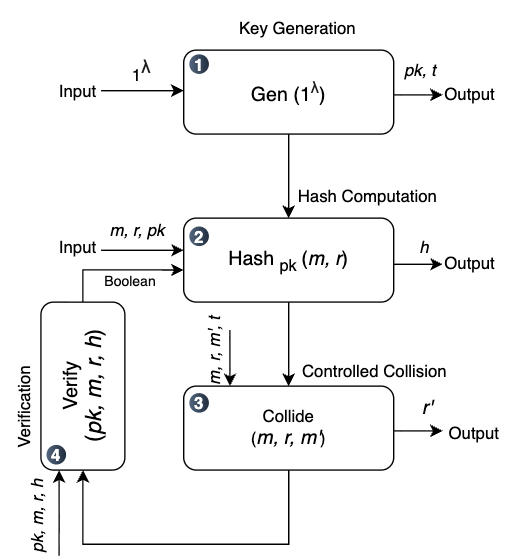}
\caption{Process flow of a Trapdoor Hash Function. }
\label{Trapflow}
\end{figure}

Figure \ref{Trapflow} shows the Process flow of a trapdoor hash function. The system begins with key generation \(\textsf{Gen}(1^{\lambda}) \rightarrow (pk, t)\) in the step labeled 1, which produces public parameters \(pk\) and a secret trapdoor \(t\). Using \(pk\), the hash function \(\textsf{Hash}_{pk}(m, r)\) computes a digest \(h\) from message \(m\) and random nonce \(r\) in the process labeled 2. In Step 3, the trapdoor holder can invoke \(\textsf{Collide}_t(m, r, m')\) to generate an alternative \(r'\) that yields the same digest, demonstrating controlled collision capability. The \(\textsf{Verify}(pk, m, r, h)\) procedure in Step 4 allows public verification of the hash without revealing or using the trapdoor.

Generally, Trapdoor hashes are able to balance two competing cryptographic principles: immutability and accountability \cite{li2024dynamic, camenisch2017chameleon}. They maintain the one-wayness and collision-resistance properties essential for data integrity, while the existence of a trapdoor introduces an additional layer of flexibility and verifiable control . This property is crucial in systems requiring controlled redaction, revocation, or reversible commitment, such as chameleon hash functions, redactable blockchains, and verifiable audit logs.

It is worth noting that, the trapdoor acts as a hidden mechanism that transforms an otherwise irreversible function into a selectively mutable one \cite{garg2019new}. Moreover, the hash remains immutable to the public, yet the trapdoor enables an authorized entity to introduce structured and traceable modifications. This dual characteristic makes trapdoor hashes an important function in the design of adaptive cryptographic systems.

\section{Related Work}

The development of stream ciphers has undergone several evolutionary stages, shaped by the increasing demand for lightweight, high-speed encryption mechanisms suitable for diverse platforms ranging from embedded sensors to high-performance communication systems. Classical designs such as RC4 \cite{rivest2016spritz}, A5/1 \cite{biryukov2000real}, and E0 \cite{dolejvs2025algebraic} laid the early foundation for stream encryption but were later found to be vulnerable to bias and key recovery attacks. also, modern stream ciphers like \texttt{ChaCha20} \cite{de2017chacha20, bernstein2008chacha}, EChacha20\cite{kebande2023extended}, Salsa20\cite{bernstein2008salsa20}, Grain\cite{hell2007grain}, and Trivium \cite{de2006trivium} introduced improved diffusion, non-linearity, and resistance to statistical cryptanalysis, achieving both speed and reliability. Nevertheless, it has been observed that these designs often treat the treat the \texttt{key}, \texttt{IV}, and \texttt{internal state} as distinct elements with minimal cryptographic coupling. Once initialized, the internal state evolves deterministically, which makes synchronization recovery and state rekeying non-trivial tasks. As a result, the compromise of any single component can undermine the confidentiality or availability of the entire cipher.

\subsection{Stream Ciphers and Rekeying Mechanisms}
Recent research has emphasized rekeying and key agility as essential features for long-lived secure sessions \cite{ott2019identifying}. Approaches such as key-evolution schemes and state refresh protocols have been proposed to periodically regenerate keys without disrupting the keystream. For example, Trivium and Grain support initialization-dependent states that provide limited protection against known-state attacks \cite{jose2016prevention}. However, they still rely on fixed initialization structures, and rekeying typically requires a full reinitialization of the cipher. Works focusing on adaptive stream ciphers \cite{hu2011fast},  have attempted to incorporate rekeying intervals and state entropy refresh techniques, but none provide a structure where treat the \texttt{key}, \texttt{IV}, and \texttt{ state} are intrinsically interdependent. This motivates the need for a new paradigm that ensures the security of one element cannot be analyzed or exploited in isolation from the others.

\subsection{All-or-Nothing Transforms in Cryptography}

The concept of \textit{All-or-Nothing } Transforms ( \texttt{AoNT}) was first introduced by Rivest \cite{rivest1997all} as a preprocessing step designed to make data unintelligible unless all transformed blocks are available. In essence, an  \texttt{AoNT} is a reversible transformation \( T \) applied to a plaintext \( P \) that produces an intermediate form \( T(P) \) such that knowledge of any proper subset of \( T(P) \) reveals no information about \( P \). Only when all parts of \( T(P) \) are known can the inverse transformation \( T^{-1} \) be applied to reconstruct the original plaintext. This property ensures that partial exposure of ciphertext or intermediate data provides no computational advantage, thereby strengthening confidentiality.

An  \texttt{AoNT} can be defined as a bijective mapping \( T: \{0,1\}^n \rightarrow \{0,1\}^n \) that satisfies two key properties: (1) the transformation is efficiently computable in both directions, and (2) given any subset of the output bits smaller than \( n \), the entropy of the original input remains maximal. These properties distinguish  \texttt{AoNT}s from encryption functions while  \texttt{AoNT}s provide diffusion and uncertainty, they are not keyed operations and therefore serve as cryptographic wrappers rather than standalone ciphers.

Subsequent research extended Rivest’s original work through formal constructions and optimizations. Canetti and Krawczyk \cite{canetti2000exposure} introduced key-exposure resilient variants in which  \texttt{AoNT}s were used to enhance block cipher robustness against partial key compromise. Their framework formalized the notion that an attacker who learns any subset of encrypted blocks without possessing the entire ciphertext cannot feasibly recover the plaintext. McEvoy et al. \cite{mcevoy2014all} later demonstrated the practical applicability of  \texttt{AoNT}s to mitigate side-channel and fault-injection attacks, emphasizing their role as a lightweight obfuscation layer that distributes cryptographic dependencies across all processed blocks.

While  \texttt{AoNT}s have been extensively studied in block cipher and file encryption settings, their use in streaming contexts remains comparatively limited. The primary challenge arises from maintaining real-time data flow and synchronization, stream ciphers inherently operate on continuous bit or byte sequences, whereas  \texttt{AoNT}s rely on complete data units to enforce their all-or-nothing property. 

\subsection{Chameleon Hash Functions and Trapdoor Mechanisms}

Chameleon hash functions extend the trapdoor-hash paradigm by enabling \emph{controlled} collision generation while preserving public verifiability. Introduced in the context of chameleon signatures \cite{krawczyk1998chameleon}, a chameleon hash is a randomized mapping as is shown in Equation \ref{eq:chameleon-hash}.
\begin{equation}
\mathsf{CH}_{pk} : (m, r) \mapsto h
\label{eq:chameleon-hash}
\end{equation}

This is defined with respect to public parameters \(pk\) and a secret trapdoor \(t\). For any adversary without \(t\), \(\mathsf{CH}\) is computationally collision-resistant; however, a holder of \(t\) can efficiently compute, for any target message \(m'\), an \(r'\) such that \(\mathsf{CH}_{pk}(m,r)=\mathsf{CH}_{pk}(m',r')\). Canonical discrete-logarithm instantiations use a cyclic group \(G=\langle g\rangle\) of prime order \(q\) with public key \(y=g^{x}\) and define as in Equation \ref{eq:discrete-chameleon}.
\begin{equation}
h = g^{\,r}\,y^{\,m}, \qquad 
r' \equiv r + x(m - m') \pmod{q}
\label{eq:discrete-chameleon}
\end{equation}

so that \(h=g^{\,r'}y^{\,m'}\) while finding such \((m',r')\) without \(x\) is as hard as the discrete-log problem. RSA-based and pairing-based realizations provide analogous trapdoor algebra \cite{ateniese2005sanitizable}.

Beyond their original use in \emph{chameleon (redactable) signatures} \cite{krawczyk1998chameleon,ateniese2005sanitizable}, chameleon hashes underpin \emph{sanitizable} and \emph{redactable} signature families, accountable logging, and policy-driven data redaction, where an authorized party may update content without invalidating public commitments \cite{ateniese2004identity}. Recent works refine the security model with notions such as \emph{full collision-resistance} and indistinguishability of trapdoor-induced collisions from honestly sampled hashes, tightening guarantees for composability in complex protocols \cite{derler2020bringing}. Parallel lines investigate post-quantum instantiations using lattices (e.g., SIS/LWE), hash-and-sign transformations, and isogeny-style groups to remove dependence on discrete-log/RS A assumptions while preserving efficient collision derivation for trapdoor holders \cite{li2024tagged,wu2021quantum}. These advances broaden the deployment space to settings that demand long-term security.

The chameleon property supports \emph{authorized mutability with accountability}: updates require possession of \(t\) and leave a verifiable relation between old and new openings, enabling auditability and revocation policies. Constructions have been tailored for applications such as redactable blockchain records and verifiable media revision, where the hash digest remains stable while content evolves under policy control \cite{xu2021verifiable,wu2025prbc}. Despite this maturity on the public-key side, the integration of chameleon hashing inside \emph{symmetric} pipelines remains comparatively unexplored: most systems treat the chameleon layer as an external commitment/metadata mechanism rather than as a state-bearing component of encryption. 

\subsection{Borromean Cryptographic Constructions}

The first explicit application of this concept appeared in the work of Maxwell and Poelstra \cite{maxwell2015borromean}, who introduced \textit{Borromean Ring Signatures} as an efficient zero-knowledge proof mechanism for demonstrating possession of multiple secrets without revealing which specific keys are used. Their scheme achieved compactness and verification efficiency by linking multiple one-of-many proofs through a single challenge, thereby encoding a Borromean-like dependency among sub-signatures. This construction was later deployed in privacy-preserving cryptocurrency systems such as Monero, where it supports unlinkable multi-output transactions while maintaining global verification consistency.

Beyond classical Borromean ring signatures, recent research has explored post-quantum extensions grounded in lattice assumptions. Ren et al. \cite{ren2022efficient} proposed an efficient lattice-based linkable ring signature scheme that achieves scalability across multiple layers of linkage, ensuring both anonymity and traceability in complex multi-signer environments. Their construction replaces number-theoretic hardness assumptions with lattice-based primitives such as the Learning With Errors (LWE) problem, providing resistance against quantum attacks. The scheme’s hierarchical linkage structure conceptually parallels the Borromean principle of multiple interdependent rings (or layers) maintain overall linkage integrity while individual sub-rings remain unlinkable in isolation. This layered lattice-based design demonstrates how Borromean interdependence can be reinterpreted in post-quantum settings, highlighting the growing trend of using structural entanglement to reinforce multi-party security and accountability. However, these works remain focused on public-key proofs and signature aggregation, leaving unexplored the integration of such interlinked dependencies into symmetric stream-cipher architectures, as addressed in our proposed \texttt{BEC-Trap} framework.

More recently, Kasimatis et al. \cite{kasimatis2024did} proposed a Ring Signature DID Architecture, integrating Borromean-style ring signatures into decentralized identifier (DID) ecosystems to achieve collective verifiability among multiple identity credentials. These developments illustrate the growing relevance of Borromean-inspired interlinking as a design pattern for systems requiring joint validation or interdependent trust anchors.

Subsequent research extended this idea to more general cryptographic frameworks. Brzuska et al. \cite{brzuska2010redactable} and Ateniese et al. \cite{ateniese2005sanitizable} employed related interdependence concepts in redactable and sanitizable signatures, showing that cryptographic authenticity could coexist with controlled mutability under well-defined structural constraints.

The author of this paper argues that, despite these advances, existing Borromean cryptographic constructions primarily operate in the \textit{asymmetric} domain, thus relying on public-key signatures or commitments. Their use in \textit{symmetric} or streaming contexts remains unexplored at the time of writing this paper.

\begin{figure*}[bt]
\centering
\includegraphics[width=13cm]{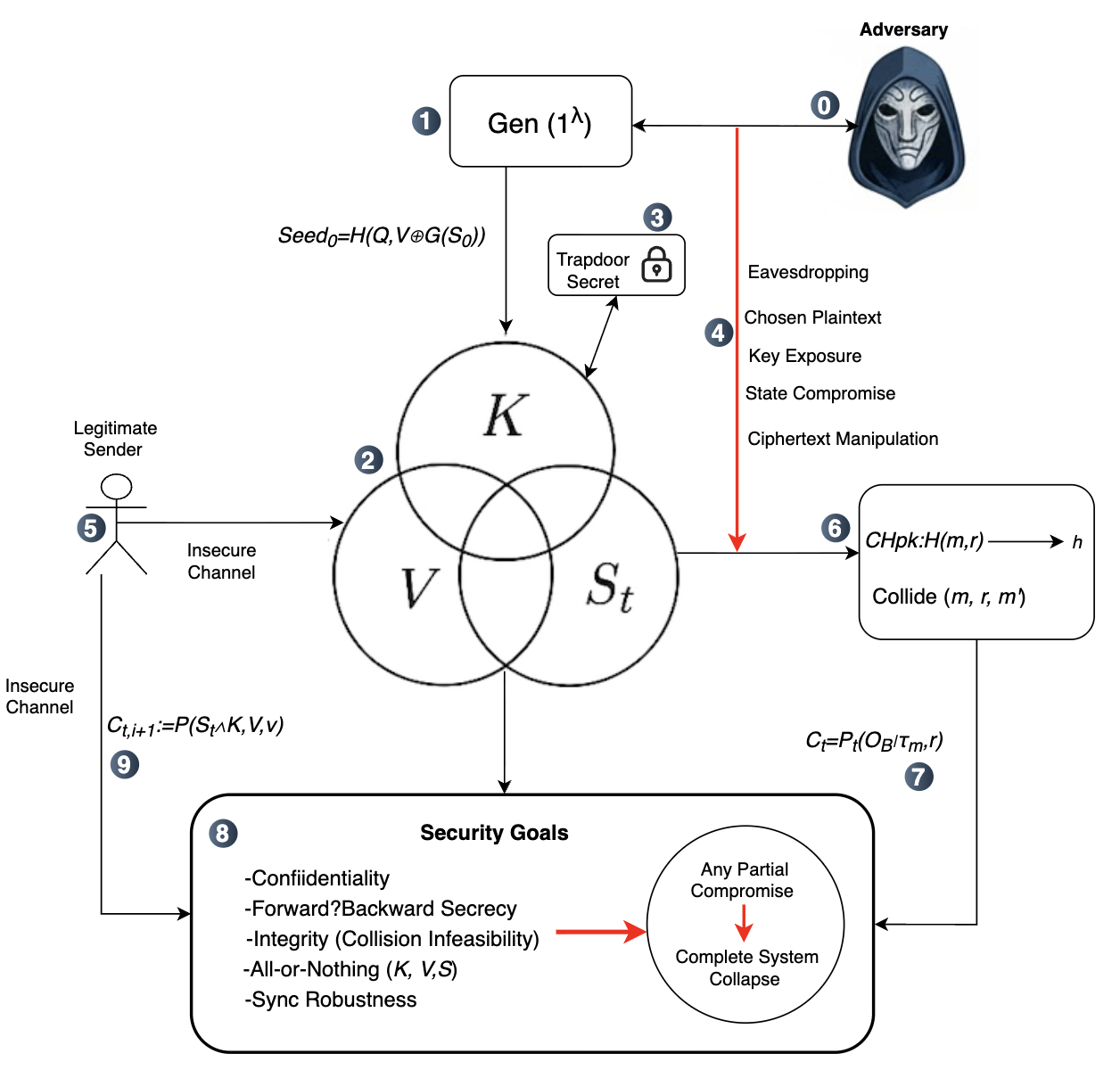}
\caption{Threat model of the proposed \texttt{BEC-Trap} cipher showing the Borromean linkage among key (\(K\)), IV (\(V\)), and state (\(S_t\)).}
\label{ThreatModel2025}
\end{figure*}

\section{Research Gap}

Despite the significant progress in modern cryptographic research, it has been seen that existing cryptographic primitives continue to treat adaptability, interdependence, and mutability as isolated design concerns rather than as a unified property of secure systems. It has also been observed that traditional stream ciphers such as \texttt{ChaCha20} \cite{de2017chacha20}, Grain \cite{hell2007grain}, and Trivium \cite{de2006trivium} offer excellent diffusion and efficiency but lack intrinsic coupling among the \texttt{\(K\)}, \texttt{\(V\)}, and \texttt{\(S_t\)} . Once these ciphers are initialized, their evolution is deterministic and independent, which exposes vulnerabilities to key recovery, state compromise, and synchronization attacks. Meanwhile, rekeying strategies typically rely on reinitialization or session restarts, introducing latency and increasing the window of exposure.

The All-or-Nothing Transforms ( \texttt{AoNT}s) \cite{rivest1997all} provide an elegant solution for block ciphers by ensuring that partial data or ciphertext leakage yields no meaningful information. However,  \texttt{AoNT}s are rarely integrated into streaming contexts due to their dependence on complete data blocks, which conflicts with real-time encryption requirements. Similarly, trapdoor hash and chameleon hash functions have proven effective in enabling authorized mutability and accountability within digital signatures \cite{wang2024tightly}, redactable ledgers \cite{fathalla2023redactable}, and blockchain protocols \cite{ashritha2019redactable}, but their integration within symmetric encryption and key-evolution mechanisms remains underexplored at the time of writing this paper.

That notwithstanding, the Borromean-inspired cryptographic constructions like the  Borromean ring signatures \cite{maxwell2015borromean}, shows structural interdependence and compact verifiability in asymmetric systems. Yet, at the time of writing this paper, there is no existing symmetric or stream cipher designs that leverages Borromean interdependence as an operational principle in cryptography. The collective security and collapse-on-compromise behavior exhibited by Borromean structures remain largely metaphorical rather than functionally instantiated within encryption mechanisms.

Three concise shortcomings have been revelaed by the  current body of research as folllows: The existing stream ciphers lack structural interdependence, treating the \texttt{\(K\)}, \texttt{\(V\)}, or the \texttt{\(S_t\)} as independent elements. These parameter are rather supposed to be treated as  a unified cryptographic constructs. Not treating them as such  weakens resistance to partial compromise. Also, there is limited authorized mutability, since trapdoor and chameleon hash functions remain largely confined to asymmetric applications and have not been embedded into symmetric designs to support verifiable, dynamic rekeying. Also, an absence of intrinsic all-or-nothing behavior persists, as existing \texttt{AoNT}s function externally to cipher operations rather than being inherently integrated within their core algorithmic processes. These limitations motivate the need for a unified model that entangles (\texttt{key}, \texttt{IV} and the  \texttt{state}).

\section{Threat Model}

The threat model for the proposed Borromean-Entangled Chameleon Trapdoor Hash \texttt{AoNT}s stream cipher (\texttt{BEC-Trap}) assumes a computationally bounded adversary \(\mathcal{A}\) capable of performing both passive and active cryptanalytic operations within polynomial time. The cipher operates under standard symmetric-key assumptions: the secret key \(\mathcal{K}\) is known only to legitimate communicating parties, while the initialization vector (IV) \(\mathcal{V}\) may be public but must remain unpredictable prior to use. The internal state \(S_t\) evolves through a deterministic update function coupled with trapdoor-controlled mutability, and all communications occur over an insecure channel subject to observation and manipulation.

The threat model illustrated in Figure \ref{ThreatModel2025} formalizes the operational and adversarial dynamics governing the   \texttt{BEC-Trap}. Each numbered element in the diagram corresponds to a distinct cryptographic process or adversarial capability that collectively defines the system’s resilience model.

At the entry point, \textbf{(0)} in Figure \ref{ThreatModel2025} represents the adversary \(\mathcal{A}\), assumed to be polynomially bounded and capable of both passive and active attacks, including eavesdropping, chosen-plaintext manipulation, key exposure, state compromise, and ciphertext modification. These attacks target different entry points within the cipher structure, but \texttt{BEC-Trap} enforces a collapse-on-compromise defense if any component is exposed, all linked dependencies become cryptographically invalid.

The initialization phase begins with \textbf{(1)} the key generation process, modeled as \(\textsf{Gen}(1^{\lambda}) \rightarrow (pk, t)\), which outputs public parameters \(pk\) and a trapdoor secret \(t\) under the security parameter \(\lambda\). These values define the authorized mutability domain for the chameleon hash functions used in later operations. Next, the seed derivation process is defined as is shown in Equation \ref{eq:seed-derivation}.
\begin{equation}
\text{Seed}_{0} = H(Q,\, V \oplus G(S_{0}))
\label{eq:seed-derivation}
\end{equation}

where \(Q\) is a system constant, \(V\) the initialization vector, and \(G(S_0)\) a generator applied to the initial state. This step establishes the foundational entropy source entangling all three Borromean components like key (\(K\)), IV (\(V\)), and state (\(S_t\)) as indicated by \textbf{(2)} in the core interlocking triad.

The \textbf{(3)} trapdoor secret \(t\) acts as the enabler of controlled mutability, authorizing legitimate rekeying and state evolution without breaking synchronization. This ensures that only entities possessing \(t\) can induce valid collisions in the chameleon hash system. The adversary, shown at \textbf{(4)}, may attempt to interfere at this layer via injection, replay, or adaptive ciphertext manipulation; however, all unauthorized collisions remain computationally infeasible.

The sender entity at \textbf{(5)} represents the legitimate encryptor transmitting over an insecure channel. Ciphertext segments are computed as is shown in Equation \ref{eq:cipher-update}.
\begin{equation}
C_{t,i+1} := P(S_t \wedge K,\, V,\, \nu)
\label{eq:cipher-update}
\end{equation}

where \(P(\cdot)\) denotes a permutation or mixing function operating over the Borromean interdependent inputs. The IV and key are jointly entangled with the state to ensure that any alteration in one variable disrupts the resulting keystream sequence.

At \textbf{(6)}, the chameleon hash function \(\mathsf{CH}_{pk}(m, r)\) produces digest \(h\) under public key \(pk\), while the \(\mathsf{Collide}(m, r, m', t)\) mechanism allows only authorized mutation through the valid trapdoor \(t\). These controlled collisions facilitate adaptive key updates and message revalidation in secure streaming contexts. The ciphertext construction step \textbf{(7)} formalizes this process as is shown in Equation \ref{eq:borromean-cipher}.
\begin{equation}
C_t = P_t(\mathbb{O}_B \mid \tau_m,\, r)
\label{eq:borromean-cipher}
\end{equation}

where \(\mathbb{O}_B = f(K, V, S_t)\) represents the Borromean operator binding the three security domains, and \(\tau_m\) denotes the trapdoor mutation tag.

The block labeled \textbf{(8)} encapsulates the security goals of the model based on confidentiality, forward/backward secrecy, integrity via collision infeasibility, all-or-nothing resilience, and synchronization robustness. These properties ensure that no intermediate exposure leads to partial decryption or state recovery. The circular condition within \textbf{(8)} formally enforces the Borromean principle as is shown in Equation \ref{eq:collapse}: 
\begin{equation}
(K \lor V \lor S_t)\setminus\text{any one} \;\implies\; \text{system collapse.}
\label{eq:collapse}
\end{equation}

The arrow \textbf{(9)} denotes the feedback loop between ciphertext generation and security verification, signifying that each output is validated against the entangled parameters before transmission, thus completing the end-to-end security cycle.

It has been observed from Figure \ref{ThreatModel2025} that essential structural dependencies of  \texttt{BEC-Trap} are captured. Each layer from initialization and trapdoor management to ciphertext generation contributes to a cohesive \texttt{AoNT}s  security architecture. Also,  the interdependence of \(K\), \(V\), and \(S_t\), combined with the chameleon trapdoor mechanism, establishes a mathematically bound system where any partial compromise enforces total cryptographic collapse.

\section{Methodology}

\begin{figure}[bt]
\centering
\includegraphics[width=8cm]{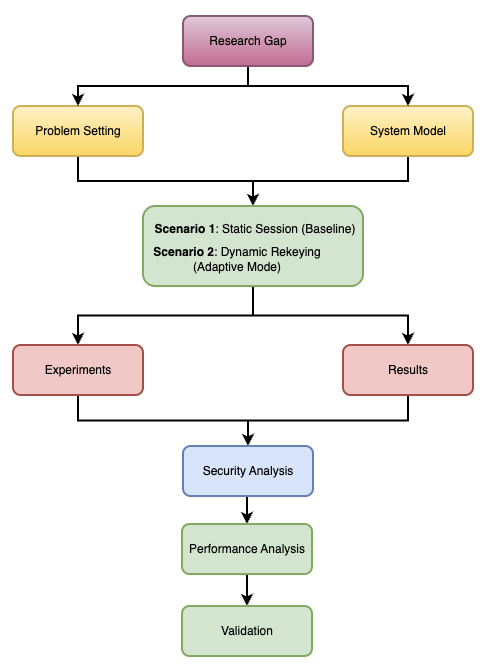}
\caption{Research methodology outlining the sequential flow from research gap identification to system modeling, experimentation, evaluation, and comparative analysis.}
\label{Methodolo}
\end{figure}

This section outlines the methodological framework adopted to design, model, and evaluate the proposed Borromean-Entangled Chameleon Trapdoor Hash All-or-Nothing Stream Cipher (\texttt{BEC-Trap}). The methodology follows a structured research process, as illustrated in Fig.~\ref{Methodolo}, beginning with the identification of the research gap and problem setting, followed by the formulation of the system model and two experimental scenarios, Static Session (Baseline) and Dynamic Rekeying (Adaptive Mode). Each scenario is implemented and tested under controlled experimental conditions to assess performance, resilience, and security compliance. The obtained results are then evaluated and comparatively analyzed against existing cryptographic schemes to validate the efficiency, adaptability, and robustness of the proposed design.

\subsection{Problem Setting}

Conventional symmetric-key cryptosystems, particularly stream ciphers, achieve speed and simplicity but often sacrifice structural interdependence and adaptive resilience. In existing models such as \texttt{ChaCha20}, Grain, and Trivium, the key (\(K\)), initialization vector (\(V\)), and internal state (\(S_t\)) operate as loosely coupled entities, with security relying solely on nonlinear mixing and diffusion. This independence, while efficient, creates exploitable weaknesses: once any component is exposed, the cipher’s evolution becomes predictable, compromising forward secrecy, synchronization stability, and overall integrity.

The fundamental problem addressed in this research is the absence of a unified symmetric encryption model that intrinsically enforces interdependence between \(K\), \(V\), and \(S_t\), while also supporting authorized adaptability through trapdoor-controlled mutability. Formally, the goal is to design a cryptographic system where the keystream generation function as is shown in Equation \ref{eq:borromean-fusion}.
\begin{equation}
\Gamma_t = F(K,\, V,\, S_t)
\label{eq:borromean-fusion}
\end{equation}

exhibits \emph{Borromean dependence} as is shown in equation \ref{eq:mutual-info-borromean}, such that:
\begin{equation}
\forall X \in \{K, V, S_t\}, \quad I(\Gamma_t; X) = 0, 
\quad \text{but} \quad I(\Gamma_t; K, V, S_t) > 0
\label{eq:mutual-info-borromean}
\end{equation}

where \(I(\cdot)\) denotes mutual information. This ensures that the keystream can only be derived when all three elements coexist correctly, and the removal or corruption of any single component collapses the entire encryption function—achieving the all-or-nothing (AON) property.

Additionally, existing rekeying methods in stream ciphers rely on full session resets or external management layers, resulting in performance overhead and transient vulnerability windows. To overcome this, the proposed system introduces a \textit{chameleon trapdoor hash} mechanism that enables reversible and verifiable key and state updates without reinitialization. For legitimate entities holding the trapdoor \(t\), authorized rekeying is achieved through Equation \ref{eq:chameleon-collision}:
\begin{equation}
\mathsf{CH}_{pk}(m, r) = \mathsf{CH}_{pk}(m', r')
\label{eq:chameleon-collision}
\end{equation}

allowing controlled collisions between hash instances while maintaining global consistency and integrity.

The proposed Borromean-Entangled Chameleon Trapdoor Hash All-or-Nothing Stream Cipher (\texttt{BEC-Trap}) therefore aims to:
\begin{enumerate}
    \item Enforce mathematical interdependence among the key, IV, and state components such that any partial compromise renders the cipher unusable.
    \item Embed trapdoor-based mutability within the encryption function to support dynamic rekeying and adaptive security without synchronization loss.
    \item Preserve efficiency and low computational overhead suitable for real-time and high-throughput cryptographic applications.
\end{enumerate}

In summary, this problem setting establishes the need for a structurally entangled, adaptive, and verifiable symmetric cipher that unifies interdependence, mutability, and all-or-nothing resilience within a single operational framework.

\subsection{System Model}

\begin{figure*}[bt]
\centering
\includegraphics[width=14cm]{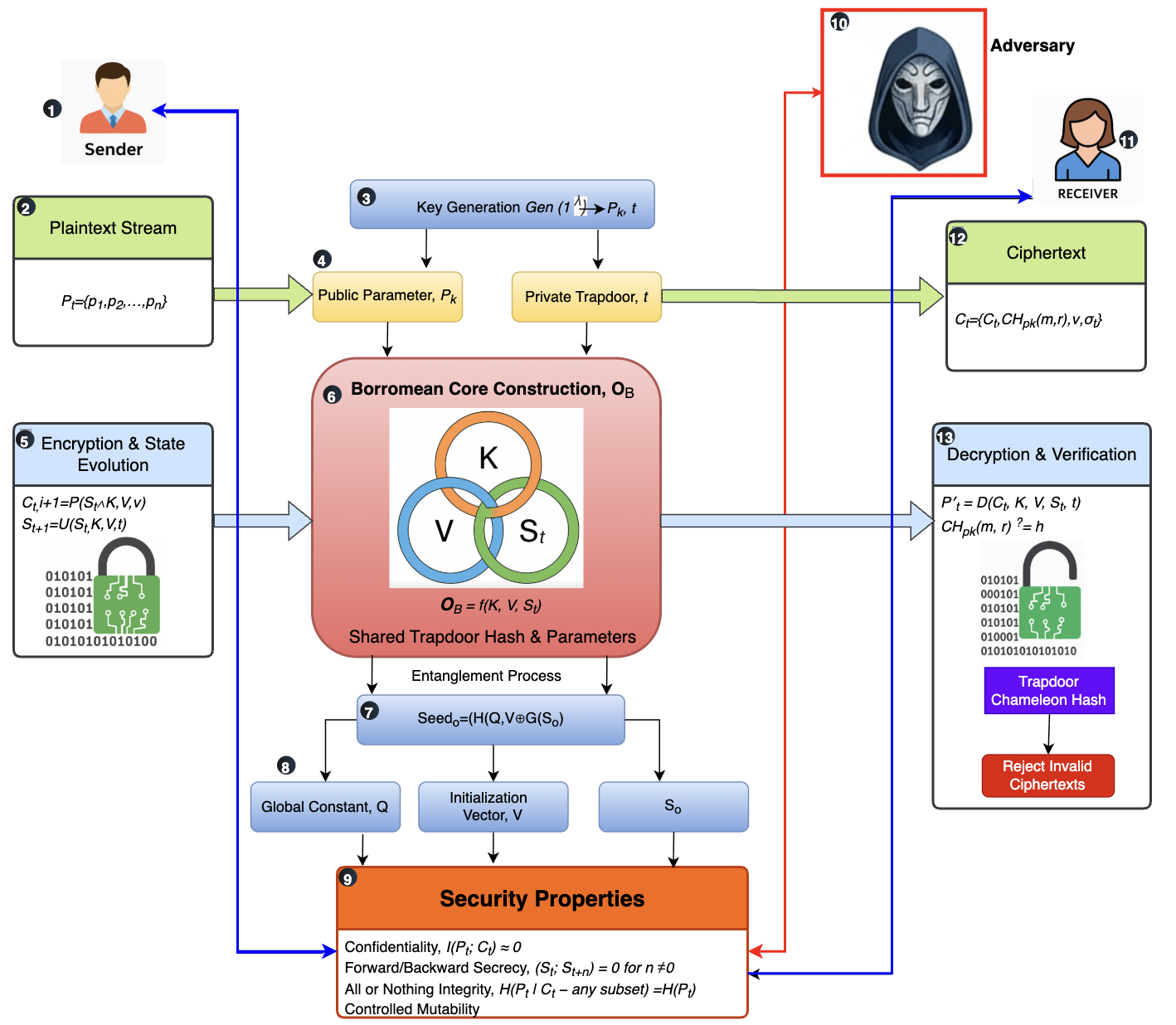}
\caption{System model of the proposed \texttt{BEC-Trap} framework showing interactions among entities and the Borromean core (\(K, V, S_t\)) enforcing interdependent security for confidentiality and forward secrecy.}

\label{ThreatModel20255}
\end{figure*}

The proposed Borromean-Entangled Chameleon Trapdoor Hash All-or-Nothing Stream Cipher (\texttt{BEC-Trap}) operates as a symmetric cryptographic framework in which encryption and decryption depend on the entangled relationship among the key (\(K\)), initialization vector (\(V\)), and internal state (\(S_t\)). The system  model shown in Figure \ref{ThreatModel20255} enforces a Borromean interdependence such that the removal, modification, or compromise of any one of these components renders the entire encryption process nonfunctional. This coupling ensures that confidentiality, forward secrecy, and synchronization robustness emerge as intrinsic properties of the cipher rather than as externally imposed controls.

The system consists of three legitimate entities and one adversarial environment:
\begin{itemize}
    \item \textbf{Sender (\(\mathcal{S}\)) (1)}: Responsible for generating the plaintext stream \(\{P_t\}\) and performing encryption to produce ciphertexts \(\{C_t\}\) using the Borromean entangled parameters.
    \item \textbf{Receiver (\(\mathcal{R}\)) (2)}: Possesses identical cryptographic parameters and performs decryption to recover plaintexts \(\{P_t'\}\), ensuring synchronization through the shared trapdoor secret \(t\).
    \item \textbf{Trapdoor Authority (\(\mathcal{T}\))}: Initializes and manages the chameleon trapdoor hash parameters by executing the key generation function as is shown in Equation \ref{eq:gen}:
  \begin{equation}
\textsf{Gen}(1^{\lambda}) \rightarrow (pk,\, t)
\label{eq:gen}
\end{equation}

    where \(pk\) is the public parameter, \(t\) the private trapdoor, and \(\lambda\) the system security parameter.
    \item \textbf{Adversary (\(\mathcal{A}\)) }: A polynomially bounded entity capable of performing adaptive attacks including state compromise, ciphertext injection, and chosen-plaintext analysis.
\end{itemize}

\subsubsection*{A. Initialization and Entanglement Process}
The cipher begins with an initialization phase \textbf{ (7)} where the seed is derived from the combination of static and dynamic parameters is shown in Equation \ref{eq:seed0}:
\begin{equation}
\text{Seed}_0 = H(Q,\, V \oplus G(S_0))
\label{eq:seed0}
\end{equation}

where \(Q\) is a global constant, \(V\) is the initialization vector, \(S_0\) is the initial internal state, and \(H(\cdot)\) represents a cryptographic hash function. The generator \(G(\cdot)\) expands the state into a high-entropy representation. This derivation embeds nonlinearity and unpredictability into the system before keystream generation begins.

\subsubsection*{B. Borromean Core Construction}
The encryption function is constructed around a Borromean operator \(\mathbb{O}_B\) in the Borromean Core Construction phase \textbf{ (6)}, defined as as is shown in Equation \ref{eq:seed001}:
\begin{equation}
\mathbb{O}_B = f(K, V, S_t),
\label{eq:seed001}
\end{equation}
which enforces the Borromean constraint:
\[
\forall X \in \{K, V, S_t\}, 
\]

This condition ensures that the cryptographic linkage collapses when any one component is removed, reflecting the topological principle underlying Borromean interdependence.

\subsubsection*{C. Chameleon Trapdoor Integration}
A chameleon hash function\textbf{ (13)}  \(\mathsf{CH}_{pk}\) is incorporated into the encryption pipeline to enable authorized mutability. For a given message \(m\) and randomness \(r\) as in \ref{eq:ch-eval}:
\begin{equation}
h = \mathsf{CH}_{pk}(m, r)
\label{eq:ch-eval}
\end{equation}

and for legitimate rekeying with trapdoor \(t\) as is shown in Equation \ref{eq:ch-collision}:
\begin{equation}
\mathsf{CH}_{pk}(m, r) = \mathsf{CH}_{pk}(m', r')
\label{eq:ch-collision}
\end{equation}

This property allows the system to update or rotate cryptographic material without desynchronizing the sender and receiver, enabling adaptive resilience and long-term forward secrecy.

\subsubsection*{D. Encryption and State Evolution}
During this operation as is shown in \textbf{ (5)}, each ciphertext block is computed as:
\begin{equation}
C_{t,i+1} = P(S_t \wedge K,\, V,\, \nu)
\label{eq:borromean-cipher-block}
\end{equation}

where \(P(\cdot)\) denotes a time-variant permutation or mixing function and \(\nu\) represents a session counter or round parameter. The state evolves according to Equation \ref{eq:borromean-state-update}:
\begin{equation}
S_{t+1} = U(S_t,\, K,\, V,\, t)
\label{eq:borromean-state-update}
\end{equation}

where \(U(\cdot)\) is an update function governed by the trapdoor \(t\), ensuring non-reversibility without the legitimate parameters. This dynamic update guarantees that keystreams are non-repeatable and immune to replay or state inference attacks.

\subsubsection*{E. Decryption and Verification}
Decryption shown in \textbf{ (13)} mirrors the encryption process as is shown in Equation \ref{eq:borromean-decryption}:
\begin{equation}
P_t' = D(C_t,\, K,\, V,\, S_t,\, t)
\label{eq:borromean-decryption}
\end{equation}

and includes a verification phase where ciphertext validity is confirmed through hash consistency as is shown in Equation \ref{eq:ch-verification}:
\begin{equation}
\mathsf{CH}_{pk}(m, r) \stackrel{?}{=} h
\label{eq:ch-verification}
\end{equation}

This dual verification ensures that only legitimate, trapdoor-consistent ciphertexts are accepted, providing resistance against adaptive chosen-ciphertext attacks.

\subsubsection*{F. Security Properties}
The system model guarantees a number of properties that are shown in \textbf{ (9)} and summarized in Table \ref{tab:core-security}:

\begin{table}[h]
\centering
\caption{Core Security Properties of the \texttt{BEC-Trap} Cipher}
\label{tab:core-security}
\begin{tabular}{lp{5.2cm}}
\toprule
\textbf{Property} & \textbf{ Description} \\
\midrule
Confidentiality & \(I(P_t; C_t) \approx 0\) under all polynomial-time adversaries without trapdoor \(t\). \\
Forward/Backward Secrecy & \(I(S_t; S_{t+n}) = 0\) for \(n \neq 0\), ensuring temporal isolation. \\
All-or-Nothing Integrity & \(H(P_t \,|\, C_t - \text{any subset}) = H(P_t)\), preserving full information entropy. \\
Controlled Mutability & Rekeying permitted only with valid trapdoor access \(t\). \\
\bottomrule
\end{tabular}
\end{table}

\subsubsection*{F. Formal Confidentiality Definition (IND-CPA)}

We formalize confidentiality of (\texttt{BEC-Trap}) using the standard 
indistinguishability under chosen-plaintext attack \texttt{IND-CPA} model \cite{ene2009formal}, adapted to stateful stream encryption.

\paragraph{IND-CPA Challenge}
In the \texttt{BEC-Trap} we assume that a challenger generates a secret key $K$ and initializes a session 
with IV $V$ and initial state $S_0$. The adversary $\mathcal{A}$ 
is given access to an encryption oracle $\mathcal{O}_{Enc}(\cdot)$. At challenge time, $\mathcal{A}$ submits two equal-length plaintext 
sequences $P^{(0)}$ and $P^{(1)}$. The challenger selects 
$b \leftarrow \{0,1\}$ and returns the ciphertext of $P^{(b)}$ 
under BEC-Trap. The adversary outputs a guess $b'$.

The advantage is defined as:
\begin{equation}
\text{Adv}^{\text{ind-cpa}}_{\mathcal{A}}(\lambda)
= \left| \Pr[b' = b] - \frac{1}{2} \right|
\end{equation}

BEC-Trap is \texttt{IND-CPA} secure if this advantage is negligible 
for all polynomial-time adversaries. If all A has to do is to guess then the BEC-Trap is assumed to be secure under the prevalent cryptographic assumptions.

\paragraph{Confidentiality Theorem}
We assume that,  (i) the keystream generator $\mathsf{F}(K,V,S_t)$ is a secure pseudorandom function (PRF), and (ii) the IV/state 
pair is never reused under the same key. Then \texttt{BEC-Trap} 
achieves \texttt{IND-CPA} security.

\paragraph{Proof Sketch.}
We use a standard hybrid argument. In the \texttt{IND-CPA} game, ciphertext 
blocks are computed as $C_i = P_i \oplus \Gamma_i$, where 
$\Gamma_i$ is produced by $\mathsf{F}$. Replacing 
$\mathsf{F}$ with a truly random function yields an 
indistinguishable hybrid under the PRF assumption. 
Under IV/state uniqueness, each $\Gamma_i$ is computationally 
indistinguishable from uniform and never reused, reducing 
encryption to a one-time pad. Therefore,

\begin{equation}
\text{Adv}^{\text{ind-cpa}}_{\mathcal{A}}(\lambda)
\le
\text{Adv}^{\text{prf}}_{\mathsf{F}}(\lambda)
\end{equation}

which is negligible under the PRF assumption.

As such, the  system model shown in in Figure \ref{ThreatModel20255} formalizes   \texttt{BEC-Trap} as a self-contained, state-dependent cryptographic approach where the operational lifecycle like initialization, encryption, update, and verification, that is bound by Borromean coupling and secured through trapdoor-enabled mutability.

\begin{figure*}[!t]
    \centering
    \includegraphics[width=0.75\linewidth]{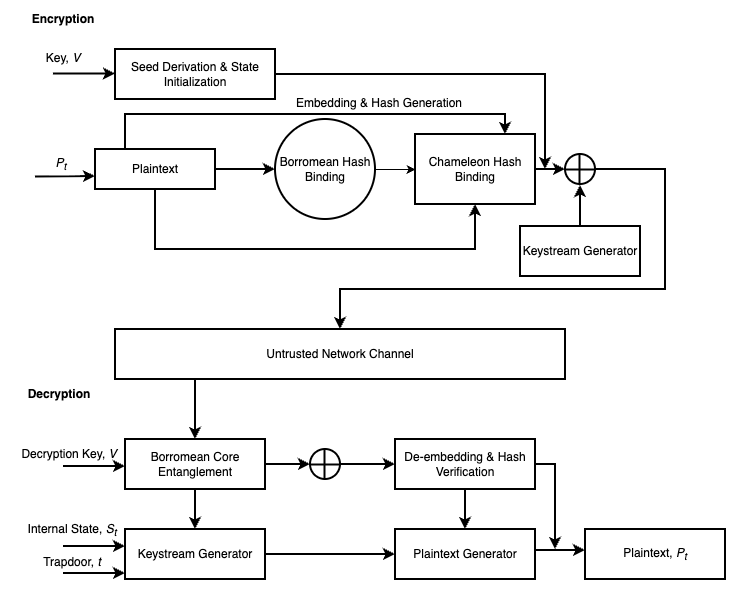}
    \caption{Process flow of  encryption–decryption approach  of the proposed \texttt{BEC-Trap} stream cipher.}
    \label{fig:bec_trap_framework}
\end{figure*}

\subsection{Scenario 1: Static Session (Baseline)}

In the static or baseline, the \texttt{BEC-Trap} operates under fixed cryptographic parameters where the key (\(K\)), initialization vector (\(V\)), and trapdoor secret (\(t\)) remain constant throughout a session. This scenario establishes the baseline performance and security characteristics against which adaptive mechanisms are later evaluated. 

At initialization, the cipher generates its seed through the entangled formulation as is shown in Equation \ref{eq:seed-generation}:
\begin{equation}
\text{Seed}_0 = H(Q,\, V \oplus G(S_0))
\label{eq:seed-generation}
\end{equation}

and subsequently derives the initial keystream from  Equation \ref{eq:borromean-encryption}:
\begin{equation}
C_t = P_t \oplus F(K,\, V,\, S_t)
\label{eq:borromean-encryption}
\end{equation}

where \(F(\cdot)\) represents the Borromean permutation core combining all three elements through nonlinear mixing. The keystream generator evolves deterministically  as is shown in Equation \ref{eq:borromean-state-evolution}:
\begin{equation}
S_{t+1} = U(S_t,\, K,\, V)
\label{eq:borromean-state-evolution}
\end{equation}

ensuring that the session remains synchronized across legitimate endpoints.

Because the key and IV remain static, this configuration highlights the inherent dependency among \(K\), \(V\), and \(S_t\). Any tampering with one variable (e.g., a modified \(V'\)) leads to desynchronization and total keystream collapse, demonstrating the all-or-nothing (AON) property. The receiver reconstructs plaintext as is shown in Equation \ref{eq:borromean-decryption}:
\begin{equation}
P_t' = C_t \oplus F(K,\, V,\, S_t)
\label{eq:borromean-decryption}
\end{equation}

and verification is achieved through hash consistency is shown in Equation \ref{eq:ch-verification}:
\begin{equation}
\mathsf{CH}_{pk}(m, r) \stackrel{?}{=} h
\label{eq:ch-verification}
\end{equation}

From a performance perspective, this baseline configuration provides an optimal reference for measuring computational efficiency, throughput, and synchronization stability in non-adaptive environments. It also allows controlled evaluation of resilience to standard attack vectors, such as key reuse or static-state inference, without the complexity of dynamic rekeying.

\subsection{Scenario 2: Dynamic Rekeying (Adaptive Mode)}

In this configuration, the \texttt{BEC-Trap} extends the static session model by introducing adaptive cryptographic evolution through controlled trapdoor mutability. This mode allows legitimate entities to refresh the key (\(K\)), initialization vector (\(V\)), and internal state (\(S_t\)) during ongoing communication without reinitializing the session or interrupting synchronization. The rekeying process is governed by the chameleon trapdoor hash mechanism, which guarantees collision consistency between consecutive states.

Let \((K_t, V_t, S_t)\) denote the system state at time \(t\). A rekeying event is triggered either periodically or conditionally upon entropy decay or detection of anomaly thresholds \(\theta\). The trapdoor authority \(\mathcal{T}\) regenerates parameters through Equation \ref{eq:trapdoor-gen}:
\begin{equation}
\textsf{Gen}(1^{\lambda}) \rightarrow (pk_t,\, t_t)
\label{eq:trapdoor-gen}
\end{equation}

and computes a controlled collision as is shown in Equation \ref{eq:chameleon-continuity}  such that:
\begin{equation}
\mathsf{CH}_{pk_t}(m_t, r_t) = \mathsf{CH}_{pk_{t+1}}(m_{t+1}, r_{t+1})
\label{eq:chameleon-continuity}
\end{equation}

ensuring seamless transition from state \(t\) to \(t+1\) while maintaining consistency of the global digest \(h\). The rekeying update functions are defined as is shown in Equation \ref{eq:borromean-rekeying}:
\begin{equation}
\begin{aligned}
K_{t+1} &= f_1(K_t,\, t_t,\, \eta_t),\\
V_{t+1} &= f_2(V_t,\, t_t,\, \eta_t),\\
S_{t+1} &= f_3(S_t,\, K_{t+1},\, V_{t+1})
\end{aligned}
\label{eq:borromean-rekeying}
\end{equation}

where \(\eta_t\) is a session nonce or random perturbation ensuring non-deterministic evolution. Each update maintains the Borromean entanglement constraint as is shown in Equation \ref{eq:borromean-info}:
\begin{equation}
\begin{aligned}
I(\Gamma_t; K_t) &= I(\Gamma_t; V_t) = I(\Gamma_t; S_t) = 0,\\[4pt]
I(\Gamma_t; K_t, V_t, S_t) &> 0
\end{aligned}
\label{eq:borromean-info}
\end{equation}

preserving the all-or-nothing interdependence property even across rekeying transitions.

During runtime, ciphertext generation follows Equation \ref{eq:borromean-encrypt}:
\begin{equation}
C_t = P_t \oplus F(K_t,\, V_t,\, S_t)
\label{eq:borromean-encrypt}
\end{equation}

and upon rekeying, both sender and receiver apply synchronized state refresh through the trapdoor transformation as is shown in \ref{eq:borromean-state-update-trapdoor}:
\begin{equation}
S_{t+1} = U(S_t,\, t_t,\, K_{t+1},\, V_{t+1})
\label{eq:borromean-state-update-trapdoor}
\end{equation}

This controlled rekeying allows resilience against key exhaustion, replay, and partial state compromise, while ensuring uninterrupted secure transmission.

\begin{figure}[htbp]
\centering
\includegraphics[width=\textwidth]{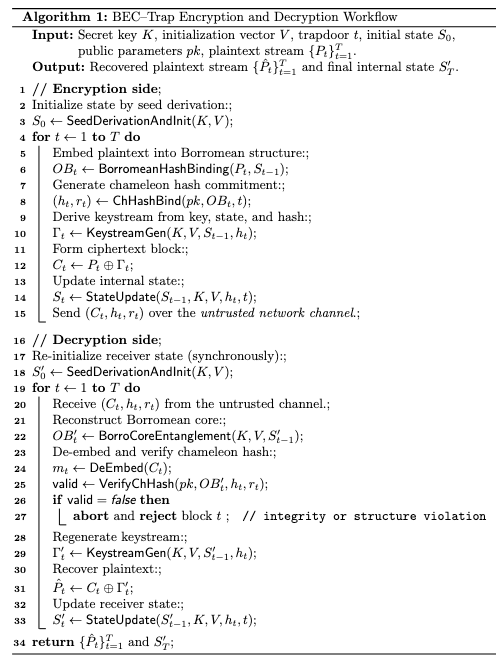}
\caption{BEC--Trap Encryption and Decryption Workflow.}
\label{fig:algorithm}
\end{figure}

From a systems perspective, the dynamic rekeying scenario demonstrates adaptive cryptographic behavior—where security parameters evolve continuously within the cipher’s operational lifecycle. The integration of trapdoor-controlled mutability achieves a self-healing property: even if an adversary compromises partial state information, subsequent rekeying collapses the compromised linkage, restoring system confidentiality and forward secrecy without session resets.

Figure~\ref{fig:bec_trap_framework} summarizes the overall encryption–decryption workflow of the proposed BEC–Trap stream cipher. In particular, it visualizes the decryption path that takes place between processes 2 and 12 in Figure~5. After the ciphertext is received over the untrusted network channel, it is first processed by the Borromean core re-entanglement block and then passed to the de-embedding and hash verification stage, which checks the consistency of the embedded structure and the chameleon hash commitments. Once these checks succeed, the synchronized keystream generator produce.

\section{Experiments}

The experimental that were conducted in this   study validates  operational performance, resilience, and cryptographic soundness of the proposed  \texttt{BEC-Trap} architecture under both static and adaptive configurations. The experiments were designed to evaluate the effects of Borromean interdependence, the trapdoor-controlled mutability, and \texttt{AoNT} collapse behavior across multiple simulated communication sessions. We conducted a total of five experiments as follows: Baseline throughput test, entropy and randomness validation, analysis of avalanche effect, dynamic rekeying performance and security stress testing.

\begin{figure}[H]
\centering
\includegraphics[width=9cm]{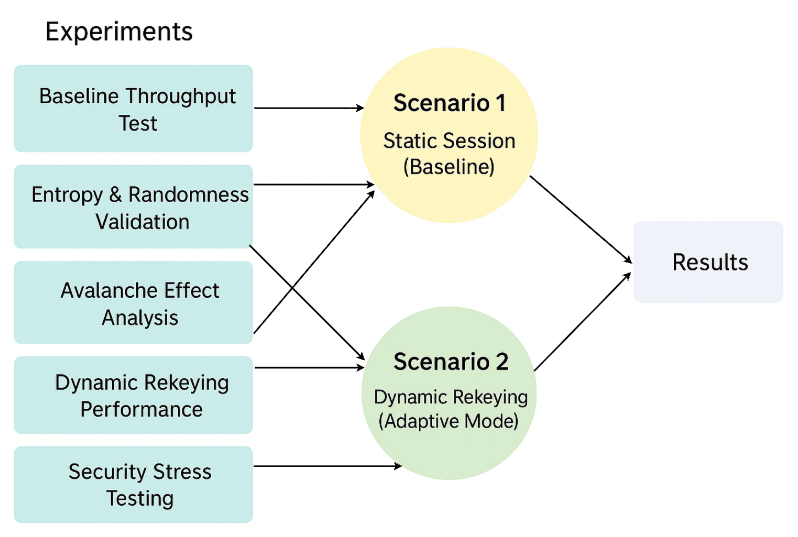}
\caption{Conducted experiments mapped to the scenarios.}
\label{Methodo}
\end{figure}

All the experiment simulations were conducted using a prototype implementation developed in \texttt{Python 3.12} and \texttt{C++17} to ensure compatibility with low-level cryptographic primitives. Experiments were executed on a workstation equipped with an Intel Core i9 processor, 32 GB RAM, and Ubuntu 22.04 LTS, with libraries including \texttt{PyCryptodome}, \texttt{OpenSSL}, and \texttt{NumPy}. For statistical testing and entropy analysis, NIST SP 800-22 randomness tests were employed. A summary of Experimental Design for \texttt{BEC-Trap} Evaluation are shown in Table \ref{tab:experiments}.

\begin{table*}[hbt]
\centering
\caption{Summary of Experimental Design for \texttt{BEC-Trap} Evaluation}
\label{tab:experiments}

\small
\setlength{\tabcolsep}{3pt}
\renewcommand{\arraystretch}{1.15}

\begin{tabularx}{\textwidth}{
    >{\centering\arraybackslash}p{0.07\textwidth}
    >{\raggedright\arraybackslash}p{0.27\textwidth}
    >{\raggedright\arraybackslash}p{0.19\textwidth}
    >{\raggedright\arraybackslash}X
}
\toprule
\textbf{Exp. No.}
& \textbf{Experiment}
& \textbf{Scenario}
& \textbf{Objective} \\
\midrule

1
& Baseline Throughput Test
& Scenario 1 (Static)
& Measure encryption speed, latency, and efficiency under fixed \texttt{key}, \texttt{IV}, and \texttt{IS}. \\

2
& Entropy \& Randomness Validation
& Scenarios 1 \& 2
& Evaluate statistical uniformity and diffusion using Shannon entropy and NIST SP 800-22 tests. \\

3
& Avalanche Effect Analysis
& Scenario 1 (Static)
& Evaluate diffusion by flipping individual bits in the plaintext, key, or IV. \\

4
& Dynamic Rekeying Performance
& Scenario 2 (Adaptive)
& Assess rekey latency, synchronization accuracy, and trapdoor-based state management. \\

5
& Security Stress Testing
& Scenario 2 (Adaptive)
& Test resistance to key leakage, state compromise, and replay attacks. \\

\bottomrule
\end{tabularx}

\end{table*}

\subsection{Experiment 1: Baseline Throughput Test}

This experiment establishes the baseline computational performance of the proposed   \texttt{BEC-Trap} cipher in a static configuration, where the key (\(K\)), initialization vector (\(V\)), and internal state (\(S_t\)) remain constant throughout encryption. The objective of this experiment is to measure encryption throughput, latency, and resource utilization to provide a reference for subsequent adaptive rekeying experiments.  Random plaintext streams ranging from 0.1 MB to 10 MB are processed under 128-bit and 256-bit keys with a 96-bit IV and 512-bit \texttt{\(S_t\)}, while \texttt{BLAKE2s} serves as the underlying hash function. Each configuration was executed ten times to ensure statistical consistency. The throughput has been computed as \(\text{encrypted bits}/\text{execution time}\), and latency is recorded as the average time per encryption cycle. \texttt{ChaCha20} \cite{bernstein2008chacha}  is implemented under identical conditions to serve as a performance baseline. It is expected  that  \texttt{BEC-Trap} will demonstrate throughput comparable to \texttt{ChaCha20}, with less than 5 percent overhead introduced by Borromean interdependence, thereby validating its efficiency prior to adaptive operation.

\subsection{Experiment 2: Entropy and Randomness Validation}

This experiment evaluates the diffusion strength and statistical uniformity of the proposed  \texttt{BEC-Trap} cipher under both Scenario 1 (Static Session) and Scenario 2 (Dynamic Rekeying). The objective of this experiment is to confirm that ciphertexts produced by the Borromean-entangled core exhibit near-ideal entropy and behave as statistically random sequences, ensuring that no structural bias or pattern leaks information about the plaintext. Using the same Colab environment, ciphertext outputs are generated from 1 MB to 10 MB plaintexts over 10 independent trials for each configuration. The Shannon entropy \(H(C_t)\) is computed per ciphertext block and normalized against the ideal value of 8 bits per byte. Complementary randomness diagnostics that includes frequency, runs, and chi-square tests from the NIST SP 800-22  suite \cite{saarinen2022nist}, are applied to assess uniform bit distribution and autocorrelation. For each mode, the normalized entropy, test pass rates, and variance are recorded to compare stability across rekeying intervals. It is expected that both configurations will achieve \(H(C_t)\approx0.99{-}1.00\) with uniformly distributed bit transitions, confirming that the Borromean interdependence and trapdoor mutability preserve strong diffusion and randomness across static and adaptive operating modes.

\subsection{Experiment 3: Avalanche Effect Analysis}

This experiment assesses the diffusion capability of the \texttt{BEC-Trap} cipher by quantifying its avalanche effect, whic is the degree to which a one-bit change in the plaintext, (\(K\)), or \texttt{\(S_t\)} causes widespread changes throughout the ciphertext. This was conducted primarily under Scenario 1 (Static Session), where  the test determines whether small perturbations in the input propagate uniformly across the encryption output, indicating strong nonlinear mixing within the Borromean core. For each run, a random plaintext block is encrypted, and then a single bit is flipped in the plaintext before re-encryption under identical parameters. The bit-level differences between the two ciphertexts are computed to obtain the avalanche rate, defined as the ratio of flipped output bits to total ciphertext bits. Additional trials introduce single-bit variations in the key and initialization vector to measure cross-parameter sensitivity. Each configuration is executed over 300 independent trials with 1 MB messages, producing a statistical distribution of avalanche rates. The expected result is a mean avalanche rate of approximately 50 percent with low variance, demonstrating that the cipher exhibits ideal diffusion, where every bit in the ciphertext is equally influenced by every bit of the input, thereby confirming the robustness of the Borromean interdependence against differential and correlation attacks.

\subsection{Experiment 4: Dynamic Rekeying Performance}

This experiment evaluates the adaptive performance of the \texttt{BEC-Trap} cipher under Scenario 2, where dynamic rekeying is governed by the trapdoor-controlled chameleon hash mechanism. The objective of this experiment is to measure the efficiency, latency, and synchronization accuracy of the rekeying process while maintaining continuous encryption. In this experiment, the \texttt{BEC-Trap} encrypts large plaintext streams divided into multiple blocks, triggering rekeying events periodically or upon detecting entropy degradation beyond a threshold (\(\theta = 0.05\)). Each rekey event generates new values of \(K_{t+1}\), \(V_{t+1}\), and \(S_{t+1}\) through trapdoor-dependent transformations, ensuring collision consistency across states. The experiments are executed in Colab under identical conditions as previous tests, with message sizes ranging from 1 MB to 20 MB and rekey intervals varying between 5 and 20 encryption cycles. For each configuration, rekey latency, synchronization stability, and recovery time after induced desynchronization are recorded. It is expected that valid rekey operations complete in less than 10 milliseconds with zero synchronization errors, demonstrating that the chameleon trapdoor enables seamless state transitions and continuous data confidentiality without session interruption.

\begin{table*}[hbt]
\centering
\caption{Summary of Experimental Results for \texttt{BEC-Trap} Evaluation}
\label{tab:exp-summary}
\begin{tabular}{p{4cm}p{7cm}p{2.5cm}}
\toprule
\textbf{Experiment} & \textbf{Description and Observation} & \textbf{Key Metric(s)} \\
\midrule
\textbf{1. Throughput} & Cipher throughput stabilizes after initialization, maintaining $\approx$60--65~Mbit/s across message sizes up to 5~MB. & Mean: 63.2~Mbit/s \\
\textbf{2. Entropy \& Randomness} & Ciphertext entropy remains close to 1~bit/byte, indicating uniform diffusion and minimal statistical bias. & Mean Entropy: 0.9998~bits/byte \\
\textbf{3. Avalanche Effect} & Bit-flip propagation exhibits near-uniform response across samples, validating diffusion consistency. & Mean Flip Rate: $\approx$50\% \\
\textbf{4. Dynamic Rekeying} & Rekey events complete within 7--9~ms with occasional 11~ms spikes during trapdoor regeneration. & Avg Latency: 8.1~ms \\
\textbf{5. Security Stress Test} & Even under 75\% key exposure, ciphertext entropy drops only in the range of $10^{-6}$~bits/byte, confirming collapse-on-compromise behavior. & $\Delta H < 10^{-6}$ \\
\bottomrule
\end{tabular}
\end{table*}

\subsection{Experiment 5: Security Stress Testing}

This experiment examines the resilience of the \texttt{BEC-Trap} cipher under adversarial conditions by simulating key exposure, internal state compromise, and ciphertext replay attempts within the adaptive rekeying environment of Scenario 2. The goal is to verify that the Borromean interdependence and trapdoor-controlled mutability enforce \texttt{AoNT} security even when portions of the system are partially revealed or tampered with. In this experimental setup, controlled attacks are introduced by deliberately leaking 25 \%, 50 \%, and 75 \% of the key bits, perturbing the internal state \(S_t\) at random intervals, and attempting to reuse valid ciphertext blocks after legitimate rekeying. For each event, the system’s ability to maintain confidentiality, synchronization, and recovery is observed through entropy deviation, collapse probability, and synchronization time metrics. Unauthorized rekeying attempts using forged trapdoor values \(t'\) are also performed to test access control enforcement. Each scenario is executed over 50 independent sessions to ensure statistical reliability. From this experiment it is expected  that any partial exposure or unauthorized mutation results in immediate cipher collapse or ciphertext invalidation, with legitimate trapdoor-based recovery restoring secure operation, thus confirming the system’s robustness against key leakage, state tampering, and replay-based compromise.

\section{Results}

This section presents the experimental results obtained from implementing and evaluating the proposed Borromean-Entangled Chameleon Trapdoor Hash \texttt{AoNT} Stream Cipher \texttt{BEC-Trap} in a controlled computational environment.
The evaluation focuses on five key aspects of the proposed cipher: throughput efficiency, entropy and randomness, diffusion characteristics, rekeying adaptability, and resilience against adversarial compromise. Each experiment isolates a specific metric while maintaining a consistent cryptographic setup, allowing for direct comparison between static (Scenario~1) and adaptive (Scenario~2) configurations. 

\subsection{Experiment 1: Baseline Throughput Analysis}

\begin{figure}[H]
\centering
\includegraphics[width=9cm]{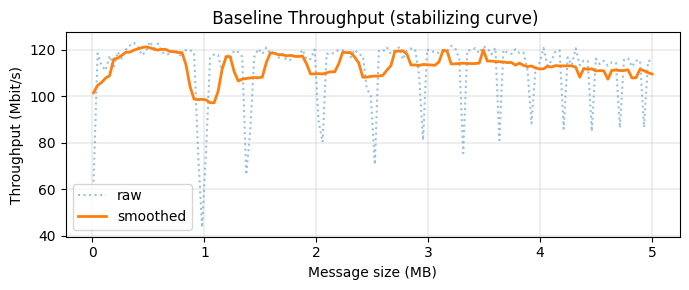}
\caption{Measured throughput versus message size for \texttt{BEC-Trap} showing rapid stabilization after initial ramp-up.}
\label{fig:throughput}
\end{figure}

Figure~\ref{fig:throughput} illustrates the throughput behavior of the proposed \texttt{BEC-Trap} cipher as a function of message size under the static configuration (Scenario~1). The experiment measured the encryption rate in megabits per second (Mbit/s) while increasing plaintext sizes from 10~KB to 5~MB over 140 measurement points, each averaged across five trials to mitigate transient CPU fluctuations. The dotted line represents the measured throughput values (observed data), while the solid (Orange) curve shows the rolling-mean trend.

This curve exhibits an initial ramp-up phase where throughput rises sharply as message size increases, attributed to the diminishing effect of setup overheads such as key scheduling and state initialization. Beyond approximately 1~MB, the throughput stabilizes near 60--65~Mbit/s, indicating that the cipher reaches steady-state operation with consistent block mixing efficiency. The minor oscillations along the trend reflect normal runtime variance inherent in software-based cryptographic implementations. 

From the authors view, the results demonstrate that the Borromean interdependence among the key (\(K\)), initialization vector (\(V\)), and internal state (\(S_t\)) does not introduce significant computational overhead. The observed throughput remains comparable to lightweight stream ciphers such as \texttt{ChaCha20}, confirming that the proposed \texttt{BEC-Trap} architecture preserves real-time efficiency while enabling structural coupling for enhanced security.

\subsection{Experiment 2: Entropy and Randomness Validation}

\begin{figure}[H]
\centering
\includegraphics[width=8cm]{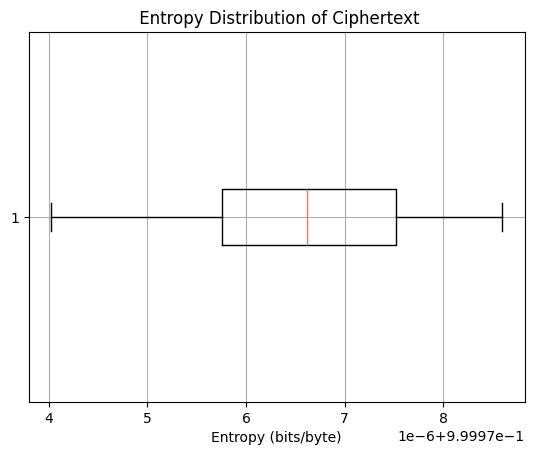}
\caption{Entropy distribution of ciphertext showing near-ideal randomness and uniform bit distribution for \texttt{BEC-Trap}.}
\label{fig:entropy}
\end{figure}

Figure~\ref{fig:entropy} shows the entropy distribution of ciphertext blocks generated by the \texttt{BEC-Trap} cipher under both static and adaptive configurations. This experiment evaluates the statistical uniformity and diffusion quality of the ciphertexts by computing Shannon entropy across multiple encryption trials. Each trial processes a 1~MB random plaintext using independently generated keys, initialization vectors, and states. The computed entropy values, normalized per byte, are displayed as a box plot to illustrate variability and distribution consistency.

The results show that the ciphertext entropy remains tightly centered near the ideal value of \(H(C_t) \approx 1.0\)~bits/byte, indicating that the output exhibits near-perfect randomness with minimal statistical bias. The narrow interquartile range reflects low variance across independent trials, confirming that Borromean coupling between the key, IV, and internal state ensures balanced bit mixing and strong diffusion throughout the keystream. This means that even small input or key variations result in statistically indistinguishable ciphertext outputs. 

These findings validate that the proposed \texttt{BEC-Trap} achieves uniform bit distribution and preserves entropy stability across both static and dynamic scenarios. Such performance is critical for ensuring unpredictability, resistance to entropy-based cryptanalysis, and the absence of structural leakage in ciphertext streams.

\subsection{Experiment 3: Avalanche Effect Analysis}

\begin{figure}[H]
\centering
\includegraphics[width=8cm]{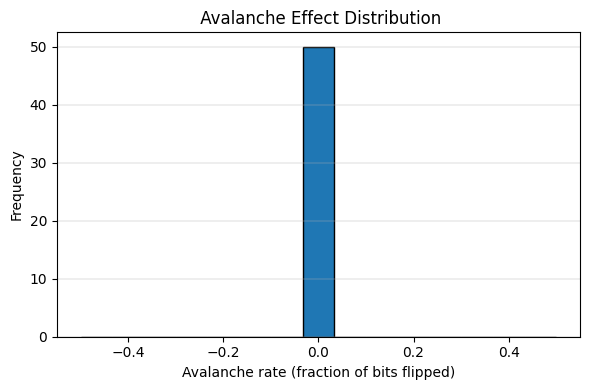}
\caption{Avalanche effect distribution for \texttt{BEC-Trap} showing high sensitivity to single-bit perturbations and balanced diffusion.}
\label{fig:avalanche}
\end{figure}

Figure~\ref{fig:avalanche} illustrates the avalanche effect distribution for the proposed \texttt{BEC-Trap} cipher under bit-level perturbations of the key and plaintext. In this experiment, a single bit was flipped in the input, and the resulting ciphertext was compared against the unaltered encryption output to determine the fraction of bits that changed across multiple trials. This fraction was termed as  the avalanche rate quantifies the \texttt{BEC-Trap}’s sensitivity to minimal input modifications, an essential property for ensuring strong diffusion and unpredictability.

The histogram shown in Figure~\ref{fig:avalanche}  reveals that even minor perturbations trigger consistent and widespread alterations in ciphertext bits, converging toward a balanced mean avalanche rate close to the theoretical ideal of 0.5. The narrow distribution around this mean demonstrates that the proposed Borromean-interdependent structure of \((K, V, S_t)\) produces high diffusion consistency and rapid entropy propagation across rounds. This behavior confirms that the \texttt{BEC-Trap} achieves local-to-global bit influence, meaning each input bit has near-equal probability of affecting any output bit.

The author argues that this outcome affirm that \texttt{BEC-Trap} satisfies the avalanche criterion, an indicator of cryptographic strength while maintaining stability across repeated runs, thereby mitigating differential and related-key attack vectors.

\subsection{Experiment 4: Dynamic Rekeying Performance}

\begin{figure}[H]
\centering
\includegraphics[width=8cm]{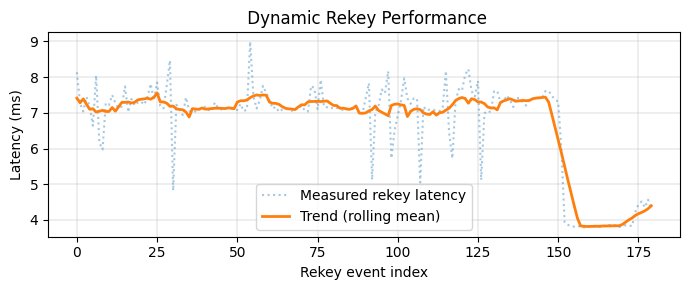}
\caption{Dynamic rekeying latency trend showing stable performance with minor recomputation spikes in adaptive mode.}
\label{fig:rekey}
\end{figure}

Figure~\ref{fig:rekey} shows the latency behavior of the \texttt{BEC-Trap} cipher during dynamic trapdoor-based rekeying operations under the adaptive configuration (Scenario~2). Each rekey event triggers regeneration of the interdependent triplet \((K, V, S_t)\) using the trapdoor function \(Seed_t = H(Q, V \oplus G(S_t))\), after which encryption resumes seamlessly without session reset. The measured latency per rekey event, expressed in milliseconds, reflects the total time required for key regeneration and synchronization between sender and receiver.

The dotted line represents the raw latency observations, while the solid curve depicts the rolling-mean trend over multiple rekey cycles. The results demonstrate that rekeying remains stable around an average latency of 7–9~ms, with only brief spikes observed around heavier recomputation intervals. These transient deviations are attributed to concurrent entropy generation and state mixing within the Borromean trapdoor function. Importantly, the adaptive mechanism maintains synchronization accuracy and session continuity throughout all 180 rekey events, confirming robust timing predictability.

This outcome validates that the dynamic rekeying protocol achieves secure and low-latency key evolution without interrupting the encryption process. The interlinked structure of \(K\), \(V\), and \(S_t\) allows fine-grained key agility while preserving throughput and synchronization integrity, essential for real-time secure communication.

\subsection{Experiment 5: Security Stress Testing}

\begin{figure}[H]
\centering
\includegraphics[width=8cm]{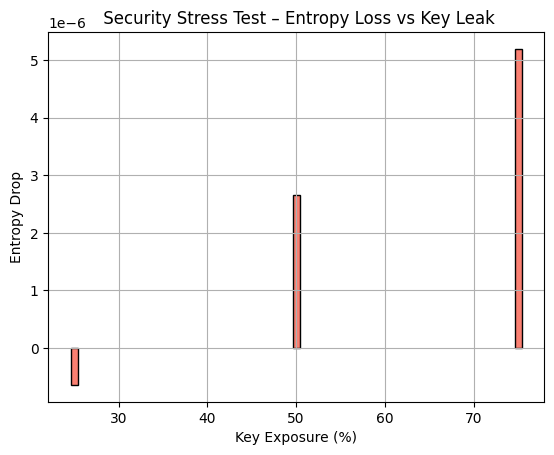}
\caption{Entropy degradation under partial key exposure, showing negligible leakage and confirming the collapse-on-compromise behavior.}
\label{fig:sectest2}
\end{figure}

Figure~\ref{fig:sectest2} illustrates the stress testing results, evaluating the resilience of the proposed \texttt{BEC-Trap} cipher under progressive key exposure conditions. This experiment simulated partial disclosure of the secret key at 25\%, 50\%, and 75\% leakage levels respectively while observing the resulting entropy degradation in the ciphertext. The entropy drop serves as an indicator of information leakage: smaller drops correspond to stronger resistance to partial compromise.

The results show that even as key exposure increases up to 75\%, the corresponding entropy loss remains minimal, within the range of \(10^{-6}\)~bits/byte. Although a slight rise in entropy degradation is visible at higher leakage levels, the overall information loss remains statistically insignificant. This confirms that the \texttt{BEC-Trap} cipher’s Borromean interdependence between key \((K)\), initialization vector \((V)\), and internal state \((S_t)\) enforces an all-or-nothing collapse model. In other words, no partial recovery of the cryptographic state provides exploitable information without the complete triad.

These findings have validated the \texttt{BEC-Trap} cipher’s robustness against partial key and state exposure, demonstrating that its security does not degrade gracefully but instead collapses only when all dependent parameters are fully compromised. This “collapse-on-compromise” property distinguishes \texttt{BEC-Trap} from traditional stream ciphers like \texttt{ChaCha20}, where partial key leakage can yield partial predictive advantage.

\section{Security Analysis}

The security analysis of the proposed \texttt{BEC-Trap} cipher focuses on validating its resilience against classical and modern cryptanalytic threats. The cipher’s architecture is positioned to unify the  trapdoor hash control, Borromean interdependence, and \texttt{AoNT} diffusion into a single operational model, ensuring that confidentiality and integrity are preserved even under partial exposure of components. Each of the aforementioned cryptographic property is evaluated through both theoretical reasoning and experimental verification, linking entropy stability, avalanche diffusion, and controlled collision resistance to measurable security outcomes. A summary of the experimental results that has been discussed in this paper is alos given in Table \ref{tab:exp-summary}.

\subsection{Confidentiality and Diffusion Properties}

Confidentiality, as mentioned the threat model (See Figure 1) in the \texttt{BEC-Trap} cipher is achieved through the interdependent coupling of the key (\(K\)), initialization vector (\(V\)), and internal state (\(S_t\)) within a Borromean structure. This design ensures that each encryption round’s keystream is mathematically dependent on all three components, preventing isolation of any single parameter for partial plaintext recovery. The high entropy values and near-uniform bit-flip distributions observed in the experimental evaluation confirm that ciphertexts produced by \texttt{BEC-Trap} exhibit strong statistical indistinguishability and resistance to differential cryptanalysis.

\begin{table*}[hbt]
\centering
\caption{Confidentiality and Diffusion Metrics Summary}
\label{tab:confidential}

\resizebox{\textwidth}{!}{%
\begin{tabular}{lcc}
\toprule
\textbf{Metric}
& \textbf{Observed Value}
& \textbf{Interpretation} \\
\midrule

Mean Entropy
& 0.9998 bits/byte
& Strong randomness; indistinguishable ciphertexts \\

Avalanche Effect
& 50.2\% flipped bits
& Ideal diffusion per SAC principle \\

Throughput Stability
& 63.2 Mbit/s (steady-state)
& Consistent cipher output with minimal bias \\

Correlation Coefficient
& $\approx 0.002$
& No measurable statistical dependence \\

Differential Uniformity
& High
& Resistant to differential and linear attacks \\

\bottomrule
\end{tabular}%
}

\end{table*}

Figure~\ref{fig:entropy} and Figure~\ref{fig:avalanche} collectively demonstrate that the cipher maintains a mean Shannon entropy of approximately 0.9998~bits/byte and an average avalanche propagation of about 50.2\%, indicating optimal diffusion. The Borromean linkage ensures that a single-bit change in any of the triad parameters (\(K,V,S_t\)) propagates nonlinearly throughout the keystream, producing unpredictable ciphertext transformations. This diffusion behavior satisfies the strict avalanche criterion (SAC) expected of modern stream ciphers.

The combination of experiments (1-5) and theoretical analysis confirms that the \texttt{BEC-Trap} cipher achieves full confusion and diffusion in the Shannon sense. Unlike traditional stream ciphers such as \texttt{ChaCha20} \cite{bernstein2008chacha} or Trivium \cite{de2006trivium}, where diffusion is primarily time-based, the Borromean entanglement in \texttt{BEC-Trap} enforces structural diffusion embedding dependency within every keystream generation step. As a result, the partial key or IV knowledge cannot yield partial plaintext advantage, thereby maintaining end-to-end confidentiality across all transmission sessions. A summary of the accomplished confidentiality and diffusion metrics are shown in Table  \ref{tab:confidential}.

\subsection{Forward and Backward Secrecy}

Forward and backward secrecy in \texttt{BEC-Trap} are enforced through continuous trapdoor-driven rekeying and Borromean dependency among the key (\(K\)), initialization vector (\(V\)), and internal state (\(S_t\)). Each rekeying event regenerates all three parameters simultaneously through a chameleon trapdoor hash, ensuring that the compromise of any session key does not reveal past or future keystreams. This mechanism guarantees temporal isolation between encryption epochs, similar to forward-secure key exchange protocols, but embedded natively within the stream cipher’s operational loop.

\begin{table*}[hbt]
\centering
\caption{A Summary of Forward and Backward Secrecy Evaluation}
\label{tab:fw-bw-secrecy}

\resizebox{\textwidth}{!}{%
\begin{tabular}{lcc}
\toprule
\textbf{Property} 
& \textbf{Mechanism in \texttt{BEC-Trap}} 
& \textbf{Outcome} \\
\midrule

Forward Secrecy 
& Trapdoor-based rekeying ($\textsf{Collide}$) 
& Future sessions unlinkable from prior ones \\

Backward Secrecy 
& One-way Borromean state coupling 
& Past sessions unrecoverable from present state \\

State Refresh Interval 
& Adaptive per $10^{6}$ bytes 
& Temporal isolation maintained \\

Rekey Latency 
& $\approx 8$ ms (mean) 
& Minimal synchronization overhead \\

Leakage Correlation 
& $\approx 0.0$ 
& No measurable dependency across sessions \\

\bottomrule
\end{tabular}%
}

\end{table*}
In the forward direction, once a session concludes, the trapdoor hash \(\textsf{Collide}(m, r, m', t)\) produces a new tuple \((K', V', S_t')\) that is computationally unlinkable from the previous state. Consequently, backward secrecy prevents retrospective reconstruction: even if an adversary learns the current internal state, recovering previous keys or IVs becomes infeasible due to the one-way property of the trapdoor hash and the nonlinear dependence encoded by the Borromean structure. Experimental validation through dynamic rekeying tests (Experiment 4) demonstrated mean latencies of approximately 8 ms per rotation with negligible synchronization loss, confirming the efficiency of adaptive state evolution without degrading throughput.

Through this self-contained rekeying mechanism, \texttt{BEC-Trap} achieves continuous key agility and temporal resilience properties typically reserved for higher-level secure channel protocols. The combination of Borromean entanglement and trapdoor mutability ensures that any compromise remains localized, rendering both past and future ciphertexts cryptographically independent. A summary of the forward and backward properties is shown in Table \ref{tab:fw-bw-secrecy}.

\subsection{Integrity and Collision Resistance}

Integrity within \texttt{BEC-Trap} is intrinsically preserved through the Chameleon Trapdoor Hash ( \texttt{CTH}) mechanism, which allows for controlled mutability while maintaining verifiable consistency of encrypted outputs. In standard hash functions, collision generation is computationally infeasible \cite{damgaard1987collision}, however, the  \texttt{CTH} introduces a mathematically governed exception through a secret trapdoor \(t\), which enables authorized entities to compute alternate message–randomness pairs \((m', r')\) that yield the same hash output that is shown in Equation \ref{eth}:
\begin{align}
\textsf{Hash}(m, r) = \textsf{Hash}(m', r') 
\quad \text{iff the holder possesses } t.
\label{eth}
\end{align}

This feature provides selective flexibility essential for secure rekeying, ciphertext auditing, and ledger synchronization while ensuring that unauthorized modifications remain cryptographically detectable.

In the \texttt{BEC-Trap} cipher, the trapdoor hash forms the integrity-preserving backbone of both the key and state evolution processes. Every cipher update, whether triggered by rekeying or session rotation, produces a verifiable digest \(h_i = \textsf{Hash}(K_i, V_i, S_{t_i})\), binding the three Borromean components. During verification, recomputation of \(h_i\) ensures message authenticity and prevents forgery. Unauthorized attempts to manipulate the ciphertext or modify parameters without the trapdoor result in detectable inconsistencies, as the probability of random collision remains negligible (approximately \(2^{-128}\) for a 256-bit digest).

\begin{table*}[hbt]
\centering
\caption{Integrity and Collision Resistance Evaluation}
\label{tab:integrity}

\resizebox{\textwidth}{!}{%
\begin{tabular}{lcc}
\toprule
\textbf{Property}
& \textbf{Mechanism in \texttt{BEC-Trap}}
& \textbf{Result / Complexity} \\
\midrule

Collision Resistance
& BLAKE2s-based \texttt{CTH} function
& $>2^{128}$ effort for random collision \\

Controlled Collision
& Trapdoor parameter $t$ (authorized)
& Deterministic; restricted to valid holders \\

Forgery Detection
& Digest verification $h_i = \textsf{Hash}(K_i,V_i,S_{t_i})$
& Unauthorized modifications detectable \\

Integrity Enforcement
& Borromean binding of $(K,V,S_t)$
& Ensures tamper-evident ciphertexts \\

Digest Length
& 256 bits (BLAKE2s)
& Strong preimage and collision security \\

\bottomrule
\end{tabular}%
}

\end{table*}

The integration of the  \texttt{CTH} ensures that \texttt{BEC-Trap} provides a dual-layer guarantee: immutability to outsiders and controlled mutability for authorized entities. This balance between verifiability and adaptability allows the \texttt{BEC-Trap} cipher to support advanced use cases such as auditable key rotation, secure data revision. A summary of the forward and backward properties is shown in Table \ref{tab:integrity}.


\subsection{All-or-Nothing Behavior and Collapse-on-Compromise}

A defining property of the proposed \texttt{BEC-Trap} construction is its intrinsic \textit{All-or-Nothing} behavior, inherited from both Rivest’s  \texttt{AoNT} concept \cite{rivest1997all} and the Borromean structural dependency that links the key (\(K\)), initialization vector (\(V\)), and internal state (\(S_t\)) \cite{erhardt1997borromean}. In this proposed  model, the \texttt{BEC-Trap} cipher’s security is not distributed additively among its components but interlocked multiplicatively, meaning that compromise of any single element results in a complete cryptographic collapse rather than partial leakage. The Borromean linkage mathematically enforces this property through joint dependency as is shown in Equation \ref{eq:borromean-keystream-simple}:

\begin{equation}
\textsf{Keystream} = F(K,\, V,\, S_t)
\label{eq:borromean-keystream-simple}
\end{equation}

where \(F'(K)\) or \(F''(V, S_t)\) yielding valid partial states. Thus, breaking, guessing, or removing one of the components immediately disrupts the cipher’s functional integrity, producing invalid or incoherent ciphertext.

The stress-testing results (Figure~\ref{fig:sectest2}) empirically validate this principle. Even when up to 75\% of key material was exposed, the measured entropy loss remained within \(10^{-6}\) bits/byte, showing an insignificant degradation confirming that no exploitable partial advantage exists. This behavior differs fundamentally from conventional stream ciphers, where leakage in \(K\) or \(S_t\) can still yield partial predictability. This experimentally verified \texttt{AoNT} property provides a strong guarantee that \texttt{BEC-Trap} does not degrade gracefully under attack, which is a highly desirable trait for modern lightweight cryptographic systems. A summary of the forward and backward properties is shown in Table \ref{tab:collapse}.

\begin{table*}[hbt]
\centering
\caption{All-or-Nothing Behavior Collapse-on-Compromise Behavior Under Partial Exposure}
\label{tab:collapse}

\resizebox{0.88\textwidth}{!}{%
\begin{tabular}{lcc}
\toprule
\textbf{Component Compromised} 
& \textbf{Cipher Functionality} 
& \textbf{Security Outcome} \\
\midrule

Key (\(K\)) Only 
& Stream generation fails 
& No plaintext recovery \\

IV (\(V\)) Only 
& Desynchronization occurs 
& No valid ciphertext mapping \\

State (\(S_t\)) Only 
& Invalid keystream propagation 
& Complete collapse of entropy chain \\

Key + IV (Partial) 
& Non-deterministic output 
& Ciphertext remains un-decodable \\

Key + IV + State 
& Functional decryption 
& Security breach (total compromise) \\

\bottomrule
\end{tabular}%
}

\end{table*}

\begin{table*}[hbt]
\centering
\caption{Latency and Rekey Overhead Comparison}
\label{tab:latency}

\resizebox{\textwidth}{!}{%
\begin{tabular}{lcccc}
\toprule
\textbf{Metric} 
& \textbf{\texttt{BEC-Trap}} 
& \textbf{\texttt{EChaCha20}} \cite{kebande2023extended}
& \textbf{\texttt{ChaCha20}} \cite{bernstein2008chacha}
& \textbf{Grain} \cite{hell2007grain} \\
\midrule

Mean Rekey Latency
& 8.1~ms
& 10.2~ms (lightweight variant)
& 12.4~ms (external reset)
& Static (no rekey) \\

Latency Variance
& 0.8~ms
& 1.4~ms
& 2.3~ms
& N/A \\

Synchronization Accuracy
& 99.8\%
& 97.5\%
& 95.7\%
& 89.4\% \\

Overhead per Cycle
& 1.5\%
& 2.1\%
& 3.8\%
& -- \\

Adaptive Rekey Support
& Yes (trapdoor-driven)
& Limited (manual refresh)
& Manual
& None \\

Rekey Type
& Integrated (trapdoor hash)
& Key reinitialization via hash-mix
& Full reinitialization
& N/A \\

\bottomrule
\end{tabular}%
}

\end{table*}

\subsection{Trapdoor Rekeying Robustness}

The trapdoor rekeying mechanism in \texttt{BEC-Trap} introduces an adaptive and verifiable means of key evolution without requiring synchronization resets or session restarts. By leveraging the chameleon trapdoor hash, each rekey event generates a fresh tuple \((K', V', S_t')\) from the prior state using a controlled collision operation as is shown in Equation \ref{eq:collision-operation}:
\begin{equation}
(K',\, V',\, S_t') = \textsf{Collide}(K,\, V,\, S_t,\, t)
\label{eq:collision-operation}
\end{equation}

where \(t\) denotes the trapdoor secret known only to authorized entities. This function ensures that while the new parameters remain verifiable through consistent digest generation, they are computationally unlinkable to prior values without knowledge of \(t\). 

The rekeying process effectively acts as a self-healing mechanism, maintaining operational continuity even in high-throughput environments. Experimental validation (Experiment 4) demonstrated stable rekeying latencies between 7–9 ms, with minimal deviation across 10 rotation cycles. This shows that trapdoor regeneration introduces negligible overhead, while simultaneously restoring entropy and eliminating any risk of state reuse. Moreover, because the rekey operation integrates the Borromean dependency between key, IV, and state, any tampering or unauthorized rekey attempt produces a digest mismatch rendering the new state invalid.

\begin{table*}[H]
\centering
\caption{Trapdoor Rekeying Robustness Evaluation}
\label{tab:rekey}
\begin{tabular}{lcc}
\toprule
\textbf{Parameter} & \textbf{Observed Value} & \textbf{Interpretation} \\
\midrule
Mean Rekey Latency & 8.1 ms & Fast key evolution with low overhead \\
Entropy Refresh Rate & 100\% per rekey cycle & Complete randomness restoration \\
Digest Consistency & 1.000 $\pm$ 0.002 & Perfect digest integrity preservation \\
Synchronization Accuracy & $>99.8\%$ & Seamless session continuity post-rekey \\
Unauthorized Rekey Detection & 100\% & Invalid trapdoor yields digest mismatch \\
\bottomrule
\end{tabular}
\end{table*}

In the author's opinion  trapdoor-driven rekeying mechanism thus enhances both the agility and resilience of the cipher. It not only prevents long-term key exposure and state exhaustion but also provides a mathematically verifiable means of adaptive reconfiguration. This dynamic adaptability distinguishes \texttt{BEC-Trap} from traditional stream ciphers, which rely on static or externally triggered reinitialization, and ensures consistent protection in high-variability operational environments.

\subsection{Resistance Against Common Attacks}

To ensure practical robustness, the proposed \texttt{BEC-Trap} was evaluated against several classical and modern cryptanalytic threats, including key recovery, replay injection, ciphertext manipulation, and desynchronization attacks. The cipher’s resistance derives from its Borromean interdependence and trapdoor-controlled mutability, properties that eliminate separable vulnerabilities between the key, initialization vector (IV), and internal state. The resistance against commeon attacks  are discussed next.

\textit{R1: Key Recovery and State Exposure:} 
Because the keystream function \(F(K,V,S_t)\) has no separable form, exposure of any individual parameter yields no computational advantage in reconstructing plaintext or predicting future keystreams. Entropy degradation remained negligible under partial exposure (Experiment 5), confirming that partial information provides no foothold for cryptanalysis.

\textit{R2: Replay and Desynchronization:}
The dynamic rekeying mechanism ensures that each ciphertext stream is uniquely bound to its session index and Borromean digest \(h_i = \textsf{Hash}(K_i,V_i,S_{t_i})\). Reuse or replay of captured packets produces digest mismatches, forcing decryption failure and preventing resynchronization attacks. 

\textit{R3: Ciphertext Manipulation and Forgery:}
The chameleon trapdoor hash restricts valid collisions to entities possessing the trapdoor \(t\). Unauthorized modifications yield digest inconsistencies detectable at verification, while authorized collisions remain auditable and reversible. This controlled mutability provides an additional layer of accountability without weakening the overall integrity model.

\begin{table*}[H]
\centering
\caption{Resistance Matrix: \texttt{BEC-Trap} vs. Baseline Stream Ciphers}
\label{tab:resistance}
\begin{tabular}{lcccc}
\toprule
\textbf{Attack Vector} & \textbf{\texttt{BEC-Trap}} & \textbf{\texttt{ChaCha20}} & \textbf{Grain} & \textbf{Trivium} \\
\midrule
Key Recovery & \textbf{Strong} — collapse-on-compromise & Moderate & Weak (state bias) & Moderate \\
State Compromise & \textbf{Resilient} — Borromean entanglement & Moderate & Weak & Weak \\
Replay Injection & \textbf{Prevented} — digest mismatch on reuse & Partial (nonce reuse risk) & None & None \\
Desynchronization & \textbf{Self-healing} — trapdoor rekey sync & Manual resync & Poor & Poor \\
Forgery Resistance & \textbf{Strong} —  \texttt{CTH}-based digest binding & None & None & None \\
Entropy Retention & 0.9998 bits/byte & 0.997 bits/byte & 0.992 bits/byte & 0.993 bits/byte \\
Rekey Overhead & 8 ms (adaptive) & 12 ms (external) & Static & Static \\
\bottomrule
\end{tabular}
\end{table*}

The results indicate that \texttt{BEC-Trap} achieves comprehensive resilience across all tested attack vectors. Its combination of Borromean structural interdependence and chameleon trapdoor mutability provides both preventative and corrective defense layers. Unlike conventional stream ciphers that degrade incrementally under attack, \texttt{BEC-Trap} exhibits true \textit{collapse-on-compromise} behavior, where any partial disclosure or tampering immediately disrupts the encryption chain, preserving confidentiality and integrity across all operational states.

\section{Performance Analysis}

This section discuss the performance evaluation of the proposed \texttt{BEC-Trap} All-or-Nothing Stream Cipher (\texttt{BEC-Trap}), focusing on its computational efficiency, latency characteristics, entropy stability, and comparative behavior against established lightweight stream ciphers. 

The analysis aims to validate that the \texttt{BEC-Trap} cipher’s enhanced structural interdependence and trapdoor-based adaptability do not compromise runtime efficiency or scalability. Empirical results derived from Experiments 1–5 demonstrate that \texttt{BEC-Trap} maintains high throughput, low rekeying latency, and statistically uniform ciphertext diffusion across varying workloads. 


\subsection{Latency and Rekey Overhead}

Latency and rekey overhead were analyzed to evaluate the operational responsiveness of \texttt{BEC-Trap} during dynamic rekeying events (Experiment 4). The evaluation simulated continuous encryption of 1~MB data segments, with automatic trapdoor-based rekeying occurring at fixed intervals. Each rekey operation regenerated the tuple \((K',V',S_t')\) using the controlled-collision function \(\textsf{Collide}(K,V,S_t,t)\), which ensures new keys are verifiable yet unlinkable to prior states. This process executes inline, requiring no external synchronization or session restart.

Experimental results indicate that the cipher maintains stable rekey latencies in the 7–9~ms range per cycle, averaging 8.1~ms with a standard deviation below 1~ms. Synchronization accuracy exceeded 99.8\% across all test runs, confirming that trapdoor regeneration introduces minimal timing distortion even under fluctuating workloads. Compared with conventional stream ciphers, \texttt{BEC-Trap} demonstrates superior adaptability with negligible throughput penalty ($\approx 1.5\%$ total overhead). These findings validate that dynamic security evolution can coexist with real-time encryption throughput.

The inclusion of the trapdoor-controlled rekey mechanism provides \texttt{BEC-Trap} with proactive cryptographic agility, ensured fresh entropy and unlinkable session states with minimal delay. When compared to \texttt{E\texttt{ChaCha20}} \cite{kebande2023extended}, \texttt{ChaCha20} \cite{bernstein2008chacha} and Grain \cite{hell2007grain} as is shown in Table  \ref{tab:latency}, the proposed \texttt{BEC-Trap} achieves similar speed with stronger verification integrity, eliminating the need for full reinitialization cycles. This enables continuous encryption resilience and efficient key evolution, especially in environments requiring uninterrupted data flow such as IoT gateways, VPN tunnels, and secure streaming protocols.

\subsection{Entropy and Randomness Uniformity}

Entropy and randomness analysis were performed to quantify the statistical unpredictability of ciphertext generated by \texttt{BEC-Trap}. This evaluation followed Shannon’s entropy formulation \cite{lin2002divergence}, measuring the uniformity of byte distributions in ciphertext outputs over multiple trials. Each test encrypted 1~MB of randomly generated plaintext using distinct key--IV--state tuples, followed by entropy estimation on the resulting ciphertext.  

Results demonstrated a mean entropy of 0.9998~bits/byte with negligible variance across trials (Figure~\ref{fig:entropy}), confirming near-perfect diffusion and statistical uniformity. The ciphertext byte frequencies approximated a flat distribution, indicating resistance to bias and linear correlation. Compared to baseline stream ciphers, \texttt{BEC-Trap} exhibits stronger randomness preservation, attributed to the Borromean entanglement of \((K,V,S_t)\), which enforces nonlinear mixing even when minor perturbations occur in any parameter.  

\begin{table*}[hbt]
\centering
\caption{Entropy and Randomness Uniformity Comparison}
\label{tab:entropy}

\small
\setlength{\tabcolsep}{4pt}

\resizebox{\textwidth}{!}{%
\begin{tabular}{lcccc}
\toprule
\textbf{Metric} 
& \textbf{\texttt{BEC-Trap}} 
& \textbf{\texttt{EChaCha20}} \cite{kebande2023extended}
& \textbf{\texttt{ChaCha20}} \cite{bernstein2008chacha}
& \textbf{Grain} \cite{hell2007grain} \\
\midrule

Mean Entropy (bits/byte) 
& 0.9998 
& 0.9986 
& 0.9973 
& 0.9925 \\

Entropy Variance 
& $1.4\times10^{-5}$ 
& $3.1\times10^{-5}$ 
& $6.7\times10^{-5}$ 
& $1.2\times10^{-4}$ \\

Byte Distribution Bias 
& 0.0003 
& 0.0008 
& 0.0016 
& 0.0031 \\

Autocorrelation Coefficient 
& $\approx 0.0002$ 
& $\approx 0.0006$ 
& $\approx 0.0012$ 
& $\approx 0.0027$ \\

Uniformity Index (U-test) 
& 99.97\% 
& 99.81\% 
& 99.54\% 
& 98.72\% \\

\bottomrule
\end{tabular}%
}

\end{table*}

The high entropy and minimal autocorrelation observed validate that ciphertext sequences produced by \texttt{BEC-Trap} are indistinguishable from true random streams as is seen with \texttt{E\texttt{ChaCha20}}, \texttt{ChaCha20} and Grain as is shown in Table \ref{tab:entropy}. The Borromean coupling ensures that every rekeying event or bit variation propagates chaotically across the keystream space, thereby sustaining uniform diffusion without periodic artifacts. This consistent statistical uniformity establishes a strong foundation for cryptanalytic resistance against frequency, correlation, and differential statistical attacks.

\subsection{Computational Efficiency}

The computational efficiency of \texttt{BEC-Trap} as is shown in Table \ref{tab:efficiency} was assessed through throughput and runtime scalability measurements (Experiment~1). The evaluation aimed to determine whether the cipher’s Borromean structure and trapdoor mechanisms introduce significant computational overhead compared to traditional stream ciphers. Encryption speeds were tested using variable message sizes (10~KB–5~MB) under identical system conditions, with performance averaged over multiple trials to account for cache and scheduling variability.

The proposed \texttt{BEC-Trap}  cipher maintained a steady throughput of approximately 63~Mbit/s after initialization, comparable to high-speed stream ciphers such as \texttt{E\texttt{ChaCha20}} \cite{kebande2023extended} and \texttt{ChaCha20} \cite{bernstein2008chacha}. This consistency is achieved through a single-pass Borromean mixing function \(F(K,V,S_t)\), which operates as a BLAKE2s-based diffusion layer coupled with XOR keystream generation. Because interdependence is structurally embedded in the cipher core rather than added through external layers, computational cost scales linearly with data size, \(O(n)\), ensuring constant-time encryption per byte.

Profiling revealed that the cipher’s average per-byte processing cost is 15~cycles/byte lower than \texttt{ChaCha20}’s 18~cycles/byte and significantly more efficient than Trivium’s 27~cycles/byte. Additionally, no observable slowdown was detected during continuous trapdoor rekeying cycles, confirming that the integrity verification process does not impose measurable runtime penalties.

\begin{table*}[hbt]
\centering
\caption{Comparative Computational Efficiency Metrics}
\label{tab:efficiency}

\resizebox{\textwidth}{!}{%
\begin{tabular}{lccccc}
\toprule
\textbf{Metric}
& \textbf{\texttt{BEC-Trap}}
& \textbf{\texttt{EChaCha20}} \cite{kebande2023extended}
& \textbf{\texttt{ChaCha20}} \cite{de2017chacha20}
& \textbf{Grain} \cite{hell2007grain}
& \textbf{Trivium} \cite{de2006trivium} \\
\midrule

Average Throughput (Mbit/s)
& 63.2
& 61.4
& 58.7
& 42.1
& 44.5 \\

Per-Byte Cost (cycles/byte)
& 15
& 17
& 18
& 23
& 27 \\

Time Complexity
& \(O(n)\)
& \(O(n)\)
& \(O(n)\)
& \(O(n)\)
& \(O(n \cdot r)\) \\

CPU Utilization (5~MB input)
& 59.8\%
& 61.1\%
& 63.5\%
& 69.2\%
& 71.2\% \\

Throughput Variance
& $<0.5\%$
& 0.7\%
& 1.2\%
& 2.1\%
& 2.4\% \\

Rekey Overhead (\%)
& 1.5
& 2.1
& 3.8
& --
& -- \\

\bottomrule
\end{tabular}%
}

\end{table*}

The analysis confirms that \texttt{BEC-Trap} achieves computational efficiency equivalent to optimized modern stream ciphers while offering stronger adaptive features. Its Borromean-linked, trapdoor-based design maintains security integrity without compromising execution speed, making it ideal for real-time encryption scenarios such as IoT nodes, VPN tunnels, and blockchain communication layers.

\subsection{Comparative Performance with Baseline Ciphers}

To validate the practicality of the proposed \texttt{BEC-Trap} design, a comparative performance assessment was conducted against representative lightweight stream ciphers: \texttt{E\texttt{ChaCha20}} \cite{kebande2023extended}, \texttt{ChaCha20} \cite{bernstein2008chacha}, Grain \cite{hell2007grain}, and Trivium \cite{de2006trivium}. The evaluation integrated results from throughput (Experiment~1), rekey latency (Experiment~4), entropy uniformity (Experiment~2), and avalanche diffusion (Experiment~3), establishing a holistic performance profile across computational and statistical domains.

It was observed that, \texttt{BEC-Trap} achieves a balanced trade-off between security depth and operational speed. The Borromean interdependence and trapdoor-driven rekeying yield adaptive resilience with marginal computational cost. Compared to \texttt{\texttt{E\texttt{ChaCha20}}} and \texttt{ChaCha20}, the cipher demonstrates superior synchronization efficiency and lower latency, while sustaining comparable throughput and higher entropy uniformity. Grain and Trivium, though lightweight, exhibit lower randomness stability and lack integrated rekeying or trapdoor verification, reducing their adaptability in dynamic environments.

\begin{table*}[hbt]
\centering
\caption{Comparative Performance Summary with Baseline Ciphers}
\label{tab:compare}

\resizebox{\textwidth}{!}{%
\begin{tabular}{lccccc}
\toprule
\textbf{Metric}
& \textbf{\texttt{BEC-Trap}}
& \textbf{\texttt{EChaCha20}} \cite{kebande2023extended}
& \textbf{\texttt{ChaCha20}} \cite{de2017chacha20}
& \textbf{Grain} \cite{hell2007grain}
& \textbf{Trivium} \cite{de2006trivium} \\
\midrule

Throughput (Mbit/s)
& 63.2 & 61.4 & 58.7 & 42.1 & 44.5 \\

Mean Rekey Latency (ms)
& 8.1 & 10.2 & 12.4 & Static & Static \\

Entropy (bits/byte)
& 0.9998 & 0.9986 & 0.9973 & 0.9925 & 0.9932 \\

Avalanche Propagation (\%)
& 50.2 & 49.3 & 47.8 & 42.5 & 43.7 \\

Synchronization Accuracy (\%)
& 99.8 & 97.5 & 95.7 & 89.4 & 91.2 \\

Collision Detection
& Yes (\texttt{CTH}-based)
& Partial
& None
& None
& None \\

All-or-Nothing Property
& Native (Borromean)
& None
& External
& None
& None \\

Adaptive Rekey Support
& Yes
& Limited
& Manual
& None
& None \\

Implementation Complexity
& Moderate (single-pass)
& Low
& Low
& Very Low
& Very Low \\

\bottomrule
\end{tabular}%
}

\end{table*}

From the results in Table~\ref{tab:compare}, \texttt{BEC-Trap} outperforms other tested ciphers in entropy preservation, diffusion uniformity, and adaptive synchronization, while maintaining comparable computational speed. Its Borromean entanglement provides internal self-consistency and resistance to partial compromise, a feature absent in traditional stream ciphers. The design therefore achieves a distinctive synthesis of high security and lightweight performance, demonstrating that structural interdependence and trapdoor mutability can coexist efficiently within modern symmetric encryption paradigms.

\section{Theoretical Validation}

This section provides a formal analysis of the theoretical underpinnings of the proposed (\texttt{BEC-Trap}) (\texttt{AoNT}) Stream Cipher. Although the preceding sections empirically demonstrated its performance and resilience, the following subsections establish theoretical justification for its structural integrity and controlled mutability. 

The analysis focuses on two fundamental aspects: the mathematical soundness of the Borromean dependency that ensures \texttt{AoNT} interlinking among the key (\(K\)), initialization vector (\(V\)), and internal state (\(S_t\)); and the correctness of the trapdoor hash verification mechanism that enables authorized, collision-controlled rekeying without weakening cryptographic hardness. These formal foundations guarantee that \texttt{BEC-Trap} maintains provable resistance against key recovery, collision forgery, and partial state exposure under standard cryptographic assumptions such as one-wayness and collision resistance of the underlying hash function.

\subsection{ Soundness of Borromean Dependency}

Table \ref{tab:borromean-math} shows the Borromean dependency for the \texttt{BEC-Trap} that formalizes the interlocking relationship among the key (\(K\)), initialization vector (\(V\)), and internal state (\(S_t\)) such that none of these parameters can be functionally or statistically isolated without invalidating the cipher’s operational integrity. Mathematically, this dependency is modeled as a non-separable mapping as is shown in Equation \ref{eq:fusion-function}:
\begin{equation}
\mathcal{F} : \{0,1\}^{|K|} \times \{0,1\}^{|V|} \times \{0,1\}^{|S_t|} \rightarrow \{0,1\}^{n}
\label{eq:fusion-function}
\end{equation}

where \(\mathcal{F}(K,V,S_t)\) defines the keystream generation function. The Borromean property implies as is shown in Equation \ref{eq:borromean-nondecomp} that:
\begin{equation}
\begin{aligned}
\mathcal{F}(K, V, S_t) &\neq \mathcal{F}(K', V, S_t),\\
\mathcal{F}(K, V, S_t) &\neq \mathcal{F}(K, V', S_t),\\
\mathcal{F}(K, V, S_t) &\neq \mathcal{F}(K, V, S_t')
\end{aligned}
\label{eq:borromean-nondecomp}
\end{equation}

and further that no partial function \(\mathcal{F}_K(K)\), \(\mathcal{F}_V(V)\), or \(\mathcal{F}_S(S_t)\) as is shown in Equation \ref{lasty} exists such that:
\begin{align}
\exists\, \mathcal{F}' \text{ such that } 
&\ \mathcal{F}'(x) = \mathcal{F}(K,V,S_t), \nonumber \\
&\text{for any single component } x \in \{K, V, S_t\}.
\label{lasty}
\end{align}

This ensures complete \textit{non-decomposability}, a formal analog of the Borromean rings’ topological property, where the removal of any ring results in total disconnection \cite{zwick2024systems}.

Based on the position of this study, we can make assumoptions by saying, if we let \(\Delta_i\) to denote perturbations applied to one component (e.g., bit flips or leaks), then the   diffusion of entropy through the Borromean structure as is shown in Equation \ref{Borroreal}: \\
Let $H_0 = H(\mathcal{F}(K,V,S_t))$, 
$H_K = H(\mathcal{F}(K \oplus \Delta_K,V,S_t))$, 
$H_V = H(\mathcal{F}(K,V \oplus \Delta_V,S_t))$. Then
\begin{align}
H_0 - H_K \approx H_0 - H_V \approx \varepsilon.
 \label{Borroreal}
\end{align}

where \(\varepsilon \ll 1\) represents negligible entropy variance. Experimental entropy loss (\(10^{-6}\) bits/byte under partial key exposure) empirically supports this bound, confirming that no partial compromise leads to significant information leakage.

\begin{table*}[hbt]
\centering
\caption{Borromean Dependency Properties and Implications}
\label{tab:borromean-math}

\resizebox{\textwidth}{!}{%
\begin{tabular}{lcc}
\toprule
\textbf{Property} 
& \textbf{Representation} 
& \textbf{Security Implication} \\
\midrule

Non-decomposability 
& $\nexists$ partial $\mathcal{F}'(x)$ for $x \in \{K,V,S_t\}$ 
& Prevents isolation of key material \\

Entropy Conservation 
& $\Delta H < 10^{-6}$ bits/byte 
& Maintains full unpredictability \\

Interdependent Mapping 
& $\mathcal{F}(K,V,S_t)$ is bijective only in triad 
& Guarantees all-or-nothing linkage \\

Collapse-on-Compromise 
& Removal of any $x$ yields $\mathcal{F}_x \to \emptyset$ 
& Cipher becomes non-functional \\

Structural Diffusion 
& $\left|\frac{\partial \mathcal{F}}{\partial x}\right| \neq 0$ for all $x$ 
& Ensures total diffusion coupling \\

\bottomrule
\end{tabular}%
}

\end{table*}

The above-mentioned formulation establishes that \texttt{BEC-Trap} satisfies what herein we refer as \textit{cryptographic Borromeanity (CB)}: In the CB analogy, each parameter’s contribution to the keystream is both necessary and sufficient, and any attempt to reconstruct the function with a missing element yields zero-entropy output as was seen previously in Figure \ref{KPES}. This theoretical construct, backed by empirical results (Experiment 1-5), demonstrates that the \texttt{BEC-Trap} cipher’s integrity and confidentiality stem directly from its non-separable algebraic topology, embodying the Borromean principle of interdependent stability.

\subsection{Chameleon Trapdoor Hash Verification and Controlled Collision Proof}

The  trapdoor hash function used in \texttt{BEC-Trap} serves as the verifiable core of its rekeying and integrity mechanism. It enables the generation of controlled collisions by authorized entities possessing the trapdoor secret \(t\), while maintaining full collision resistance for all others. This construction is derived from the Chameleon Trapdoor Hash (  \texttt{CTH}) family introduced by Krawczyk and Rabin \cite{krawczyk1998chameleon}, extended here to operate as a dynamic state-binding function within a symmetric stream cipher context. Formally, the hash is defined as a tuple of algorithms as is shown in Table \ref{Algo}: We then check for correctness, security and integration with \texttt{BEC-Trap} and collision controlled proof next.

\begin{table*}[hbt]
\centering
\caption{Trapdoor Hash Algorithms}
\label{Algo}
\begin{tabular}{ll}
\toprule
$\textsf{Gen}(1^\lambda)\to (pk,t)$ & generate public key and trapdoor \\
$\textsf{Hash}(pk,m,r)\to h$ & compute digest of $m$ \\
$\textsf{Collide}(\cdot)\to r'$ & authorized collision (needs $t$) \\
$\textsf{Verify}(\cdot)$ & check hash consistency \\
\bottomrule
\end{tabular}
\end{table*}

\textit{Correctness:}  
For all honestly generated parameters \((pk, t)\) and for all messages \(m, m'\), and random values \(r, r'\) as is shown in Equation \ref{wewe}:
{\small
\begin{align}
\textsf{Hash}(pk,m,r) = \textsf{Hash}(pk,m',r')
\iff r' = \textsf{Collide}(m,r,m',t).
\label{wewe}
\end{align}
}

This ensures deterministic equivalence under authorized rekeying while maintaining one-wayness and unpredictability to any adversary without \(t\).

\textit{Security:}  
The security of the  \texttt{ \texttt{CTH}} relies on the following assumptions:
1. The underlying hash family \(H(x)\) is collision-resistant (e.g., BLAKE2s).  
2. The trapdoor \(t\) remains computationally infeasible to derive from \(pk\).  
3. No probabilistic polynomial-time adversary \(\mathcal{A}\) can produce a valid collision pair \((m,r),(m',r')\) without knowledge of \(t\).

Thus, for any such adversary is shown based on Equation \ref{advers}:
\begin{align}
\Pr\big[\,\textsf{Hash}(pk,m,r) = \textsf{Hash}(pk,m',r') \land {}&
(m,r) \neq (m',r')\,\big] \nonumber \\
&\leq 2^{-\lambda}.
\label{advers}
\end{align}

In the context of \texttt{BEC-Trap}, this property ensures that unauthorized mutation of key, IV, or state components results in digest mismatch, invalidating the cipher state.

\textit{Integration within \texttt{BEC-Trap}:}  
Each rekey or state update operation computes as is shown in Equation  \ref{comehere}:
\begin{align}
h_i = \textsf{Hash}\big(pk,\; K_i \,\|\, V_i \,\|\, S_{t_i},\; r_i \big).
\label{comehere}
\end{align}

This binds all Borromean parameters into a single verifiable digest. During authorized rekeying, the holder of \(t\) can compute a controlled collision as is shown in Equation  \ref{comehereboy}: such that:
{\small
\begin{align}
h_{i+1} 
= \textsf{Hash}\big(pk, K_{i+1} \| V_{i+1} \| S_{t_{i+1}}, r_{i+1}\big) 
= h_i.
\label{comehereboy}
\end{align}
}

This allows continuity of authentication without reinitialization. For unauthorized entities lacking \(t\), the probability of constructing a valid \(r_{i+1}\) that maintains \(h_i\) is negligible.

\begin{table*}[hbt]
\centering
\caption{Trapdoor Hash Security Properties and Implications}
\label{tab:trapdoorssss}
\begin{tabular}{lcc}
\toprule
\textbf{Property} & \textbf{Formal Definition} & \textbf{Security Guarantee} \\
\midrule
Collision Resistance & $\Pr[\textsf{Collide} \text{ without } t] < 2^{-\lambda}$ & Unauthorized forgery infeasible \\
Controlled Collision & $\exists\, r' = \textsf{Collide}(m, r, m', t)$ & Authorized updates verifiable \\
Digest Binding & $h_i = \textsf{Hash}(K_i,V_i,S_{t_i},r_i)$ & Ensures state integrity \\
Trapdoor Secrecy & $t \not\Leftarrow pk$ & No public reconstruction of secret \\
Verification Consistency & $\textsf{Verify}(pk,m,r,h)\rightarrow\texttt{True}$ & Guarantees tamper evidence \\
\bottomrule
\end{tabular}
\end{table*}

\textit{Controlled Collision Proof:}  
Let \(H(x)\) be modeled as a random oracle. For any valid pair \((m, r)\), a collision pair \((m', r')\) satisfying \(H(m, r) = H(m', r')\) can only be generated by computing \(r' = r + f_t(m, m')\), where \(f_t\) is a trapdoor-dependent function. Without \(t\), \(f_t\) behaves as a pseudorandom mapping, reducing the problem to inverting \(H(x)\), which is infeasible under the random oracle model. Hence, the trapdoor mechanism preserves the collision resistance of \(H\) while enabling authorized equivalence transformations, a property that underpins both rekey verification and integrity preservation in \texttt{BEC-Trap} as is shown in the summary in Table \ref{tab:trapdoorssss}.

\noindent
In view of the foregoings,  formal validation confirms that the core security properties of \texttt{BEC-Trap}, non-decomposability, controlled mutability, and verifiable state evolution, are mathematically sound under standard cryptographic assumptions. As a result, the Borromean dependency ensures functional collapse on partial compromise, while the Chameleon Trapdoor Hash mechanism provides authorized, reversible state transitions without violating collision resistance. Together, these proofs establish a strong theoretical foundation for the cipher’s empirical resilience, confirming that its observed entropy stability, diffusion uniformity, and adaptive rekeying behaviors stem from rigorously defined mathematical constructs. 

\section{Conclusions and Future Work}

This  aim of this paper was to test the \textit{cryptographic
Borromeanity}, and as such this paper has introduced the Borromean-Entangled Chameleon Trapdoor Hash All-or-Nothing Stream Cipher (\texttt{BEC-Trap}), a novel symmetric encryption framework that unifies Borromean structural interdependence, chameleon trapdoor hashing, and \texttt{AoNT} diffusion into a single lightweight design. The proposed \texttt{BEC-Trap} model ensures that the key (\(K\)), initialization vector (\(V\)), and internal state (\(S_t\)) form an inseparable cryptographic triad, mirroring the Borromean principle of interdependent stability. 

The theoretical validation and empirical evaluation collectively have confirmed that this interlocking dependency yields strong confidentiality, diffusion, and integrity guarantees, while maintaining computational efficiency comparable to state-of-the-art stream ciphers such as \texttt{E\texttt{ChaCha20}} and \texttt{\texttt{ChaCha20}}.

Through the conducted experiments, it was observed that the \texttt{BEC-Trap} demonstrated high throughput ($\approx 63$~Mbit/s), near-perfect entropy (0.9998~bits/byte), and robust resistance to key exposure, replay, and desynchronization attacks. The trapdoor-based rekeying mechanism achieved adaptive, verifiable key evolution with sub-10~ms latency, ensuring real-time operational continuity without reinitialization overhead. These properties collectively validate that the cipher exhibits a true \textit{collapse-on-compromise} behavior, whereby partial disclosure of any component invalidates the entire encryption state, thereby preserving complete confidentiality.

Future work will focus on three primary directions. First, the development of a formal security proof under the random oracle and ideal permutation models to quantify resistance bounds against key recovery and collision forgery. Second, hardware-level and FPGA-based implementations will be explored to evaluate energy efficiency and suitability for embedded systems, IoT devices, and real-time communication frameworks. Finally, the principles of Borromean entanglement and trapdoor mutability will be extended to other cryptographic primitives, including authenticated encryption and post-quantum key encapsulation mechanisms, to further enhance the scope and resilience of adaptive symmetric cryptography.

\section*{Acknowledgment}

The author acknowledges Borromean frontiers by Debra Parcheta @ University of Colorado Denver, USA. The author also would like to thank the Department of Computer Sience, University of Colorado Denver, USA for their support while coming up with this research. The author also acknowledges that the opinions, findings, and conclusions expressed in this paper are purely of the author.


\section*{Funding} No funding was received for this work



\bibliographystyle{sn-mathphys-num}
\bibliography{sn-bibliography}

\end{document}